\documentclass[12pt]{article}
 \immediate\openin0=hyperref.sty
 \ifeof0\immediate\closein0\else\immediate\closein0\usepackage[hyperindex]{hyperref}\fi
 \let\oldensuremath=\ensuremath\renewcommand\ensuremath{\relax\oldensuremath}

 \newcommand\units[1]{\ensuremath{\mathbin{}{\rm#1}}}
 \newcommand\MeV{\units{MeV}}
 \newcommand\GeV{\units{GeV}}

\newcommand{\be}{\begin{equation}}
\newcommand{\ee}{\end{equation}}
\newcommand{\ba}{\begin{eqnarray}}
\newcommand{\ea}{\end{eqnarray}}
\newcommand{\Lambdabar}{\overline{\Lambda}}
\newcommand{\MSbar}{{\overline{MS}}}
\newcommand{\gsim}{\ensuremath{\mathrel{\raise2pt\hbox to 8pt{%
                        \raise -5pt\hbox{$\sim$}\hss{$>$}}}}}
\newcommand{\bsbd}{$B_s - B_d$ }
\newcommand{\sss}{$2S-1S$ }
\newcommand{\pminuss}{$1P-1S$ }
\newcommand{\hyperf}{$B^\ast-B$ }
\newcommand{\Lb}{$\Lambda - \overline{B}$ }
\newcommand{\sigl}{$\overline{\Sigma}- \Lambda$ }
\newcommand{\shyp}{$\Sigma^\ast - \Sigma$ }
 \newif\ifchi\newif\iflat\newif\ifMeV
 \newif\iflight\newif\ifinter\newif\ifstrange
 \newif\ifK\newif\ifKstar\newif\ifphi
 \chifalse\lattrue\MeVtrue \lighttrue\intertrue\strangetrue
 \Ktrue\Kstartrue\phifalse

 \newcount\colcount\newcount\allcolcount
 \newcommand\setupcounts{%
    \colcount=0
    \iflight  \advance\colcount by 1\fi
    \ifinter  \ifK    \advance\colcount by 1\fi
              \ifKstar\advance\colcount by 1\fi
              \ifphi  \advance\colcount by 1\fi\fi
    \ifstrange\ifK    \advance\colcount by 1\fi
              \ifKstar\advance\colcount by 1\fi
              \ifphi  \advance\colcount by 1\fi\fi
    \allcolcount=1
    \ifchi\advance\allcolcount by 1\fi
    \iflat\advance\allcolcount by \colcount\fi
    \ifMeV\advance\allcolcount by \colcount\fi
    \ifcase\colcount\begingroup\errhelp{can't switch off all columns}%
                               \errmessage{colcount is 0}\endgroup\fi
 }
 \newcommand\selectcols[7]{%
    \iflight#1\fi\ifinter\ifK#2\fi\ifKstar#3\fi\ifphi#4\fi\fi
    \ifstrange\ifK#5\fi\ifKstar#6\fi\ifphi#7\fi\fi}
 \newcommand\putrow[4]{#1\ifchi#2\fi\iflat#3\fi\ifMeV#4\fi}
\input{epsf}
\begin{document}


\title{ \vspace{-1cm}\begin{flushright}
{\small OHSTPY-HEP-T-98-018 \\
UTCCP-P-77 \\
 LAUR-99-723 \\ }
\end{flushright}
{
\vspace{1cm}
Heavy-light Mesons and Baryons with $b$ quarks 
}}

\author{
{\bf A. Ali Khan$^a$, T. Bhattacharya$^b$, S. Collins$^c$,
C.T.H. Davies$^{c}$,}\\ {\bf R. Gupta$^b$, C. Morningstar$^d$, 
J. Shigemitsu$^e$, J. Sloan$^f$\thanks{Present address: Spatial 
Technologies, Boulder, CO, USA.}}\\[.6cm]
\small $^a$ Center for Computational Physics, University of Tsukuba, \\ 
\small Tsukuba, Ibaraki 305-8577, Japan.\\[.2cm]
\small $^b$Los Alamos National Laboratory,\\ 
\small Los Alamos, NM 87545, USA.\\[.2cm]
\small $^c$Department of Physics \& Astronomy, University of Glasgow, \\
\small Glasgow, UK G12 8QQ. \\[.2cm]
\small $^d$Physics Department, Florida International University, \\
\small Miami, FL 33199, USA. \\[.2cm]
\small $^e$Physics Department, The Ohio State University,\\ 
\small Columbus, OH 43210, USA.\\[.2cm]
\small $^f$Physics Department, University of Kentucky,\\
\small Lexington, KY 40506, USA.
\\ }

\date{\today}

\maketitle

\begin{abstract}
\noindent
We present lattice results for the spectrum of mesons containing one
heavy quark and of baryons containing one or two heavy quarks. The
calculation is done in the quenched approximation using the NRQCD
formalism for the heavy quark.  We analyze the dependence of the mass
splittings on both the heavy and the light quark masses.  Meson
$P$-state fine structure and baryon hyperfine splittings are resolved
for the first time.  We fix the $b$ quark mass using both $M_B$ and
$M_{\Lambda_b}$, and our best estimate is $m_b^\MSbar(m_b^\MSbar) =
4.35(10)({}^{-3}_{+2})(20)$ GeV.  The spectrum, obtained by
interpolation to $m_b$, is compared with the experimental data.
\vspace{.1in}

\end{abstract}

\section{Introduction}
\label{sec:intro}

The spectrum and decays of hadrons containing $b$ quarks will be
measured in precision experiments at the $B$ factories.  It is
therefore important to calculate the spectrum expected from QCD, both
as a test of the theory and to predict the masses of states not yet
observed. This paper reports on results of a lattice calculation of the
heavy-light spectrum using the non-relativistic formulation of QCD
(NRQCD) for heavy quarks \cite{cornell}, and the tadpole-improved
clover action for light quarks. This approach allows us to have better
control over discretization errors in both the heavy and the light
quark sectors.

Lattice QCD allows us to investigate the dependence of the meson and
baryon mass splittings on the heavy and light quark masses.  For this
purpose we simulate three values of light quark masses in the range
$0.8m_s - 1.3m_s$, and six values of heavy quark masses in the range
$3 - 20$ GeV. The NRQCD formalism is ideally suited to study such a
wide range of heavy quark masses at $1/a = 1.92$ GeV, the lattice
spacing we use.  For the light quarks we use the tadpole-improved
clover action which has discretization errors of $O(\alpha_s a)$ and
these are expected to be small at this lattice spacing. These
improvements make it possible to perform reliable comparisons with
both the experimental $b$ spectrum and expectations based on Heavy
Quark Symmetry.

The phenomenological interest in the decay rates of hadrons containing
$b$ quarks stems from the important role they play in the
determination of the Cabibbo-Kobayashi-Maskawa matrix elements. Two
quantities that are used as input in the analyses of experimental data
are $m_b^\MSbar(m_b^\MSbar)$ and the decay constants $f_B$ and
$f_{B_s}$. Here  we shall present results for the $b$ quark mass, while 
the calculation of decay constants has already been reported in a 
companion paper \cite{fBpaper}. 

This paper is organized as follows.  In Sec.~\ref{sec:expt} we
briefly review the experimental situation and provide a justification
for the NRQCD approach to heavy quarks.  The parameters used in the
simulations are given in Sec.~\ref{sec:simulation}.
Sec.~\ref{sec:mb} describes the determination of the $b$ quark
mass.  Our results on the heavy-light meson spectrum are presented in
Sec.~\ref{sec:mesons} along with a discussion of the
spin-independent and spin-dependent mass splittings.  Baryons
containing one heavy and two light quarks are discussed in
Sec.~\ref{sec:HLLbaryons}. In Sec.~\ref{sec:HHLbaryons} we give
a brief description of our results on baryons containing two
(degenerate) heavy quarks and one light quark.  This is followed by a
determination of HQET parameters in Sec.~\ref{sec:HQET}.  Finally,
we summarize our main conclusions in Sec.~\ref{sec:conclusions}.

\section{Phenomenological Background}
\label{sec:expt}

\begin{figure}[tbp]  
\begin{center}
\epsfysize=3in
\centerline{\epsfbox{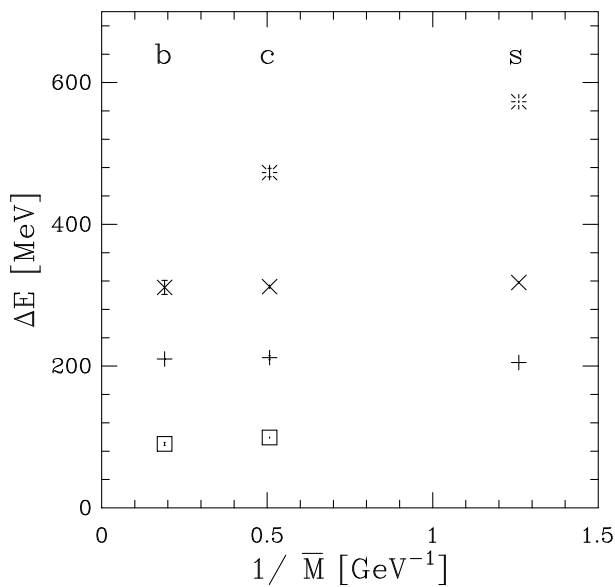 }}
\vspace{-1cm}
\end{center}
\caption{ Experimental spin-independent mass splittings for hadrons
with one heavy quark ($h = b,\ c,$ or $s$) as a function of the
spin-averaged meson mass $\overline{M} \equiv (M_H + 3 M_{H^\ast})/4$ where 
$H$ denotes a generic heavy meson.
Squares denote the $B_s-B_d$ and the $D_s-D_d$ splitting.  Pluses
stand for the spin-averaged $\Sigma-\Lambda$ splitting (we have used
the DELPHI measurement of $\Sigma_b$ \protect \cite{lep2}).  The
splitting between the $\Lambda$ and the spin-averaged $S$ state meson
is denoted by crosses. Bursts denote the spin-averaged $P-S$
splitting.
}
\label{fig:overview1}
\end{figure}

\begin{figure}[tbp] 
\epsfysize=3in
\begin{center}
\centerline{\epsfbox{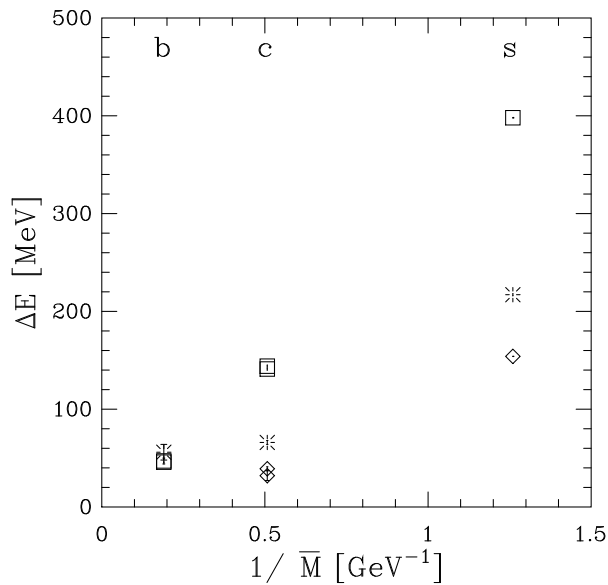 }}
\vspace{-1cm}
\end{center}
\caption{Experimental spin-dependent mass splittings for hadrons with
one heavy quark. $\overline M$ is defined in Fig.~\protect\ref{fig:overview1}. 
Squares denote the $S$ state hyperfine splitting for
$B$, $B_s$, $D$, $D_s$, and $K$ mesons.  Diamonds denote the splitting
between $P$ states with $j_l$ = 3/2. These are known only for $D$,
$D_s$, and $K$ mesons. The $\Sigma^\ast-\Sigma$ splitting for baryons
is denoted by bursts.  (For a discussion of the possibility that some of the $\Sigma_c$ and
$\Sigma_b$ baryons have been misidentified experimentally see 
Ref.~\protect\cite{falk97}).}
\label{fig:overview2}
\end{figure}

The NRQCD approach for simulating $b$ quarks is justified because the
typical velocity of the heavy quark is small, $v/c \sim
O(\Lambda_{QCD} / M) \sim 0.05 - 0.1$.  This is corroborated by the
experimental observation that all splittings are much smaller than the
masses, and the hadron masses are dominated by the heavy quark mass.
Thus a very natural picture of the heavy-light system is a ``hydrogen
atom'' composed of the light degrees of freedom bound in the
background of an almost static color source. Within this model one can
distinguish between spin-independent splittings in the spectrum
dominated by radial and/or orbital excitations of the light quark, and
spin-dependent ones dominated by the spin-flip energy of the heavy
quark. These two types of splittings have distinct behavior as a
function of the heavy quark mass. Spin-independent splittings survive
the infinite heavy quark mass limit whereas the spin-dependent ones do
not.

The experimental data plotted in Fig.~\ref{fig:overview1} show that
the spin-independent splittings are often insensitive to the mass of
the heavy quark.  In fact one finds in many cases that the
insensitivity persists down to the strange quark mass.  Spin-dependent
splittings, on the other hand, are found to increase with the inverse
heavy quark mass as shown in Fig.~\ref{fig:overview2}. An 
analysis with a phenomenologically determined potential is in
agreement with these results, however there is considerable
uncertainty in how to model the light degrees of freedom
(see~\cite{potentialinvert} and references therein).  
Simulations of lattice
QCD using a non-relativistic formulation for heavy quarks provide
estimates without resort to modeling.  

The NRQCD formulation has been discussed in~\cite{cornell,upsilon}. It
has been very successful in the study of heavy
quarkonia~\cite{upsilon}, and we apply it to predict the heavy-light
spectrum here.  Results using alternate formulations, static heavy
quarks or standard (Wilson or clover) discretization of the Dirac operator
mostly extrapolated from the charm region, can be found in
\cite{review,chrism,UKQCDbaryons,peterboyle,jimsimone} and we shall
compare against them at appropriate places.

\section{Simulation Parameters}
\label{sec:simulation}

\begin{table}[tbp]
\begin{center}
\begin{tabular}{|c||c|}
\hline
\multicolumn{2}{|c|}{Heavy quark parameters} \\
\hline
$\begin{array}{c}
aM^0 \\n \end{array}$ & $\begin{array}{c|c|c|c|c|c}
         1.6 & $2.0$ & $2.7$ & $4.0$ & $7.0$ & $10.0$ \\
          2   &  2    &  2    &  1    &  1    &   1 \\
        \end{array}$ \\
\hline
\multicolumn{2}{|c|}{Light quark parameters} \\
\hline
$\begin{array}{c}
\kappa \\ am_\pi\end{array}$ & $\begin{array}{l|l|l}
0.1369 & 0.1375 & 0.13808 \\
0.423(7) & 0.362(11) & 0.298(4) \\
\end{array}$ \\
\hline
\end{tabular}
\end{center}
\caption{A summary of the heavy and light quark mass parameters,
$aM^0$ and $\kappa$, used in the simulation and the resulting mass of
pions composed of degenerate light quarks. We also list the values of
the stability parameter $n$ used in the heavy quark
evolution~\protect\cite{fBpaper}. }
\label{tab:params}
\end{table}

\begin{table}[thb]
\begin{center}
\setlength{\tabcolsep}{0.25cm}
\begin{tabular}{|ll|l|}
\hline
\multicolumn{2}{|c|}{state}&\multicolumn{1}{c|}{operator} \\
\hline
$^1S_0$    &     & $\bar u \;h$ \\
$^3S_1$    &     & $\bar u \;\vec \sigma \;h$ \\
$^1P_1$    &     & $\bar u \;\vec \Delta \;h$ \\
$^3P_0$    &     & $\bar u \;\vec \sigma \cdot \vec\Delta \;h$ \\
$^3P_1$    &     & $\bar u \;\vec \sigma \times \vec\Delta \;h$ \\
$^3P_2$    & (T) & $\bar u \;(\sigma_i \Delta_{j} + \sigma_j \Delta_{i}) \;h$ \\
$^3P_2$    & (E) & $\bar u \;(\sigma_i \Delta_i - \sigma_j \Delta_j) \;h$ \\
\hline
$\Lambda$  & ($s_z = +1/2$) & $\bar u^c d \; h_\uparrow$ \\
$\Sigma$   & ($s_z = +1/2$) & $(\bar u^c \sigma_z d \; h_\uparrow 
                                - \sqrt{2} \bar u^c \sigma_+ d \; h_\downarrow)
                              /\sqrt{3}$ \\
$\Sigma^*$ & ($s_z = +3/2$) & $\bar u^c \sigma_+ d \; h_\uparrow$ \\
$\Sigma^*$ & ($s_z = +1/2$) & $(\sqrt{2} \bar u^c \sigma_z d \; h_\uparrow 
                                +  \bar u^c \sigma_+ d \; h_\downarrow)
                              /\sqrt{3}$ \\
\hline
\end{tabular}
\end{center}
\caption{The operators used to study the various states. $h$ stands for the
two-component heavy quark spinor, $u$ and $d$ for the upper two
components of two flavors of light quark spinors. $\vec \Delta$ is the
ordinary derivative in the Coulomb gauge.  The symbols $h_\uparrow$
and $h_\downarrow$ stand for the $s_z = +\frac12$ and $-\frac12$
components of $h$ respectively.  The baryon operators for $s_z < 0$
are constructed from the corresponding $s_z > 0$ operators by
interchanging $\sigma_+ \leftrightarrow \sigma_-$ and $\uparrow
\leftrightarrow \downarrow$.  The $\Xi$ baryons are obtained by replacing 
one of the light flavors in $\Sigma$ by an $s$, and the $\Omega$ by replacing both light quarks 
by $ss$. For the heavy-heavy-light baryons, the
operators are identical except $u$ and $d$ are to be interpreted as
two flavors of heavy quarks and $h$ as the light or $s$ quark. The ${}^3P_2$ states 
decompose, under the cubic group, into  two representations labeled T and E.
Our $j = 2$ $P$ states are
spin-averaged over both lattice representations: ${}^3P_2 =
[3\mathbin{}{}^3P_2(T) +  2\mathbin{}{}^3P_2(E)]/5$. 
}

\label{tab:ops}
\end{table}

The statistical sample consists of the same 102 quenched
configurations, at $\beta = 6.0$ with lattice size $16^3\times 48$, as
used in our study of decay constants~\cite{fBpaper}.  The NRQCD
action, the evolution equation for calculating the heavy quark
propagator, the method used for setting the lattice scale, and the
fixing of light and strange quark masses are also the same. The list
of quark masses used in our simulation are reproduced in
Table~\ref{tab:params}, and the operators used to study the various
states, are given in Table~\ref{tab:ops}. 

We estimate that the significant sources of systematic errors in this 
calculation
are finite volume, finite lattice spacing, quenching, uncertainties in
determining $a$, fixing the strange quark mass and perturbative
corrections. For a lattice size of $\approx 1.6$~fm, finite volume
effects are not expected to be significant for the lower lying $S$
state mesons. However, there are indications that the wave functions
for $P$ states and the baryons are more extended~\cite{chrism} and
finite size effects in these states should therefore be larger. We
cannot comment on this as we have results on only one lattice volume.
The $O(\alpha_s a)$ error associated with the tadpole improved clover
light fermions is expected to be a few percent at this
$\beta$~\cite{hartmut}. A detailed study of the scaling behavior
of the heavy-light spectrum is discussed in Ref.~\cite{scalepaper}.
Quenching errors remain unknown. However,
since the $B$ spectrum is dominated by the light quark degrees of
freedom, we expect that using light spectroscopic quantities to fix
$a$ compensates for part of this uncertainty.

The central value of lattice scale we
use is $1/a = 1.92(7)$ GeV as obtained from $M_\rho$. 
To estimate the
systematic error in this we repeat our bootstrap
analyses with $1/a = 1.8$ and $2$ GeV as discussed in \cite{fBpaper}.
We obtain $\kappa_l = 0.13917(9)$, corresponding to the light quark
mass $m_l = 1/2(m_u+m_d)$, by linearly extrapolating
$M_\pi^2/M_\rho^2$ to $137^2/770^2$. We cannot resolve a
curvature in the light quark mass dependence, and do not assign a systematic 
error in $\kappa_l$. To determine the strange quark
mass, we use three different methods. By fixing the ratio
$M_K^2/M_\pi^2$ to its physical value, we obtain $\kappa_s =
0.13755(13)$.  Using the ratios $M_{K^\ast}/M_\rho$ and
$M_\phi/M_\rho$, gives $\kappa_s = 0.13719(25)$ and $0.13717(25)$
respectively. Since the latter two agree within errors, we only give
the results using $M_K$ and $M_{K^\ast}$. For our final results, we use
$\kappa_s$ from $M_K$, and determine the systematic error using
$\kappa_s$ from $K^\ast$. 

In our final results, the first error we quote comes from a 
bootstrap analysis using $a^{-1} = 1.92(7)$ GeV, the second from the
scale uncertainty, and where applicable, the third from the
uncertainty in the strange quark mass. We comment on the uncertainty
due to using 1-loop perturbative expressions and in fixing the $b$ 
quark mass below.

A summary of some of the important features of the raw lattice data
are as follows.  (i)~The data at $aM^0=7.0$ and $10.0$ are not as
reliable as that for $aM^0 \le 4.0$ (there are no clear plateaux in the
effective mass plots). They are, therefore, used only in the
estimation of HQET parameters, where we have chosen states and operators
with the best signal.
(ii) The calculation of $P$ state correlation functions has been
done for only $aM^0 = 1.6, 2.0, 2.7$.


Lastly, we fix the bare $b$ quark mass $aM_b^0$ as follows.  In NRQCD
and the static theory, $E_{\rm sim}$, the rate of exponential fall-off
of the heavy-light meson correlators, is not the meson mass, but is
related to it by the shift,
\be 
\Delta = M_{\rm hadron} - E_{\rm sim} = Z_m M^0 - E_0 \,. 
\label{eq:Deltadef}
\ee 
Here $Z_m$ is the renormalization constant connecting the bare quark
mass to the pole mass, and $E_0$ is the shift in the energy of the
quark.  As discussed in more detail in Ref.~\cite{fBpaper}, we employ
three different methods to calculate the meson mass: (i) $M_{\rm kin} $
extracted directly from the dispersion relation of the heavy-light
meson; (ii) $M_{\rm pert}$ obtained by evaluating the mass shift $\Delta$
perturbatively; and (iii) $M^\prime$ using the $\Delta$ obtained from
the dispersion relation of the heavy-heavy meson at the same $aM^0$
value.  The perturbative results for $Z_m$, $E_0$ and $\Delta$ are
given in Table~\ref{tab:pert}.\footnote{The perturbative calculations have been
done for a slightly different discretization of $F_{\mu\nu}$, $i.e.$ a
four leaf clover rather than the two leaf version used in the
evolution equation. We expect the difference to be insignificant.} 

\begin{table}[thbp]
\begin{center}
 \setlength\tabcolsep{0.25cm}
\begin{tabular}{|r|c|l|l|l|}
\noalign{\hrule}
$aM^0$ & $n$ & $Z_m$     & $aE_0$   & $a\Delta$ \\
\noalign{\hrule}
1.6    &  2  & 1.18  & 0.23 & 1.64  \\
2.0    &  2  & 1.14  & 0.28 & 2.02  \\
2.7    &  2  & 1.09  & 0.27 & 2.68  \\
4.0    &  1  & 1.05  & 0.27 & 3.90  \\
7.0    &  1  & 1.00  & 0.28 & 6.74  \\
10.0   &  1  & 0.98  & 0.28 & 9.54  \\
\noalign{\hrule}
\end{tabular}
\end{center}
\caption{The stability parameter $n$ and the 1-loop perturbative 
estimates of the mass renormalization
constant $Z_m$, the zero point shift of the heavy quark energy
$E_0$, and the mass shift $\Delta = Z_m M^0 - E_0$ using the 
$q^\ast$ calculated with the Hornbostel-Lepage 
procedure~\protect\cite{lephorn}.
Errors associated with numerical integration of the 1-loop expressions 
are insignificant compared to other systematic errors.}
\label{tab:pert}
\end{table}
%


In the perturbative analyses, we use $\alpha_s=\alpha_P$ defined in
Ref.~\cite{alpha1}.  The relevant scale $q^\ast$ at which to evaluate
the running coupling $\alpha_P$ is chosen separately for each process
using an extension~\cite{lephorn} of the Lepage-Mackenzie
scale-setting prescription~\cite{lepmac}.  The choice of scale
advocated in the original Lepage-Mackenzie scheme eliminates the
$O(\alpha_s^2)$ correction in the bubble summation approximation.
This procedure can fail, however, when the one-loop contribution
becomes small.  Hornbostel and Lepage~\cite{lephorn} have recently
extended the method to overcome this difficulty by taking into account
higher-order terms in the bubble summation approximation.  Their
extension reduces to the original Lepage-Mackenzie prescription when
the one-loop term is not small due to large cancelling contributions.

The perturbative series for $Z_m$ has an infra-red renormalon
ambiguity~\cite{RENORMALONBB}, which is typically characterized by an
uncertainty of $O(\Lambda_{QCD}/M)$.  Since this is comparable to the
entire $O(\alpha_s)$ correction, we shall use the latter as the
estimate of the perturbative error in the determination of $M_{\rm
pole}$.

\begin{table}
\begin{center}
 \setlength\tabcolsep{0.15cm}
\begin{tabular}{|l|l|l|l|l|l|l|l|l|}
\hline
\multicolumn{1}{|c}{} &
\multicolumn{4}{|c|}{$\kappa_{l}$} &
\multicolumn{4}{|c|}{$\kappa_{s}$} \\
\hline
\multicolumn{1}{|c|}{$aM^0$} &
\multicolumn{1}{|c}{$aE_{\rm sim}$} &
\multicolumn{1}{|c}{$aM_{\rm kin}$} &
\multicolumn{1}{|c|}{$aM'$} &
\multicolumn{1}{|c}{$aM_{\rm pert}$} &
\multicolumn{1}{|c}{$aE_{\rm sim}$} &
\multicolumn{1}{|c}{$aM_{\rm kin}$} &
\multicolumn{1}{|c}{$aM'$} &
\multicolumn{1}{|c|}{$aM_{\rm pert}$} \\
\hline
1.6 &0.427(7)& 2.16(13) &2.08(3) &2.07(4) &0.474(4)& 2.21(9)  &2.13(3) &2.11(2)\\  
2.0 &0.443(8)& 2.57(18) &2.46(4) &2.46(2) &0.490(4)& 2.63(13) &2.50(4) &2.51(1)\\  
2.7 &0.459(7)& 3.30(30) &3.15(7) &3.14(2) &0.504(4)& 3.35(21) &3.20(7) &3.18(1)\\  
4.0 &0.468(8)& 4.76(65) &4.46(11)&4.37(7) &0.513(5)& 4.74(43) &4.50(11)&4.41(7)\\ 
7.0 &0.469(9)& 8.9(24)  &        &7.21(21)&0.516(5)& 8.5(15)  &        &7.26(21)\\ 
10. &0.471(9)& 15(7)    &        &10.01(35)&0.515(5)&13(4)    &        &10.1(3)\\ 
\hline       
\end{tabular}
\end{center}
\caption{$E_{\rm sim}$ and pseudoscalar meson masses in lattice
units extrapolated/interpolated to $\kappa_l$ and $\kappa_s$. Meson masses have
been calculated from the heavy-light dispersion relation ($M_{\rm
kin}$), using $\Delta$ from heavy-heavy spectroscopy ($M'$), and from
perturbation theory ($M_{\rm pert}$).}
\label{tab:masses}
\end{table}

All three methods for estimating the $B$ meson mass give compatible
results for $aM^0 \le 4$ as shown in Table~\ref{tab:masses}. These
estimates differ slightly from those in Ref.~\cite{fBpaper} due to a
reanalysis of the data and different choice of $q^*$.
Unfortunately, the most direct method, using the heavy-light
dispersion relation, has large errors.  The method using $\Delta$
extracted from heavy-heavy mesons is more accurate for $aM^0
\le 4.0$. For $aM^0 = 7.0$ and $10.0$, heavy-heavy mesons have large
discretization errors as these are governed by $pa \sim \alpha_s M a$,
so the corresponding data for $\Delta$ are not reliable.  To
summarize, the best estimate is $aM_b^0 = 2.31(12)$ obtained by
matching $M'$ to the pseudoscalar meson mass, $M_B = 5279$ MeV.  Using
$M_{\rm kin}$ instead of $M'$ gives a consistent determination,
$aM_b^0 = 2.21(22)$, though with larger errors.

A comparison of the three similar ways of determining $M_b^0$ using the
$\Lambda_h$ baryon mass is presented in Table~\ref{tab:M2Lambdab}.
Here, and in the following, we use the symbol $\Lambda_h$ to represent
a heavy-light-light $\Lambda$ baryon with $h$ labeling the heavy quark.
Again, we find that 
the difference between the three methods are significant only for
$aM^0 = 7.0$ and $10.0$.  Therefore, we determine $M_b^0$ by linearly
interpolating the data at the lightest three $M^0$ values. The result
is $aM_b^0 = 2.5(6)$ using $M_{\Lambda_b} = 5624 $ MeV. This is consistent
with  the estimate from the meson sector; however, since it has 
much larger errors we do not consider it further.

The final issue in fixing $aM_b^0$
is related to the fact that our calculation fails to
reproduce the experimental hyperfine splitting between the $B$ and the
$B^\ast$, as discussed in Sec.~\ref{ss:BstarB}. Thus, it could be argued that
determining $aM_b^0$  from the spin-averaged 
$1S$ mass $(m_B + 3 m_{B^\ast})/4 = 5313$ MeV should give a better
estimate.  We find that $aM_b^0 = 2.32(12)$  obtained by matching $M'$
to the spin-averaged mass is in complete agreement with the value obtained
from $m_B$. Henceforth, we shall use the value $aM_b^0=2.32(12)$ for the
$b$ quark mass.

\begin{table}[htb]
\begin{center}
 \setlength\tabcolsep{0.15cm}
\begin{tabular}{|l|l|l|l|l|l|l|l|l|}
\hline
\multicolumn{1}{|c}{} &
\multicolumn{4}{|c|}{$\kappa_{l}$} &
\multicolumn{4}{|c|}{$\kappa_{s}$} \\
\hline
\multicolumn{1}{|c|}{$aM^0$} &
\multicolumn{1}{|c}{$aE_{\rm sim}$} &
\multicolumn{1}{|c}{$aM_{\rm kin}$} &
\multicolumn{1}{|c|}{$aM'$} &
\multicolumn{1}{|c}{$aM_{\rm pert}$} &
\multicolumn{1}{|c}{$aE_{\rm sim}$} &
\multicolumn{1}{|c}{$aM_{\rm kin}$} &
\multicolumn{1}{|c}{$aM'$} &
\multicolumn{1}{|c|}{$aM_{\rm pert}$} \\
\hline
1.6 &0.626(25)&2.19(31)&2.28(4) & 2.27(5)  &0.751(14)&2.54(16)&2.41(4) & 2.39(4)\\  
2.0 &0.645(30)&2.56(40)&2.66(5) & 2.67(4)  &0.766(16)&2.93(20)&2.78(4) & 2.79(3)\\  
2.7 &0.660(37)&3.21(53)&3.35(8) & 3.34(5)  &0.777(18)&3.62(29)&3.47(7) & 3.46(3)\\  
4.0 &0.688(59)&4.7(11) &4.68(12)& 4.59(12) &0.785(28)&5.02(50)&4.77(11)& 4.69(9)\\ 
7.0 &0.702(52)&9.2(11) &        & 7.44(25) &0.783(28)&8.95(56)&        & 7.52(23)\\ 
10. &0.726(72)&16.4(27)&        & 10.2(4)  &0.783(38)&14.2(14)&        & 10.3(3)\\ 
\hline       
\end{tabular}
\end{center}
\caption{$E_{\rm sim}$ and $\Lambda_h$ baryon masses in lattice units. 
Symbols have the same meaning as in Table~\protect\ref{tab:masses}.}
\label{tab:M2Lambdab}
\end{table}

\section{Mass of the $b$ quark, $m_{\overline{MS}} (m_{\overline{MS}})$}
\label{sec:mb}

There are two steps needed to determine quark masses from lattice
calculations.  First, the bare quark masses have to be fixed by
matching the lattice spectrum to experimental data. This has been described
in Sec.~\ref{sec:simulation}. Next, one needs to calculate the
renormalization constants that relate these bare masses to the
renormalized mass in the desired continuum scheme.  The most common
scheme is $\overline{MS}$ and we shall use it here. Standard continuum
perturbation theory calculations can then be used to convert the result to
any other scheme.

We calculate the $\overline{MS}$ mass by equating the pole mass on the
lattice to that in the continuum:
\begin{equation}
m_{pole} = Z_m M_b^0 = Z_{\rm cont}(\mu) m_{\overline{MS}} (\mu) \,,
\label{Mpole}
\end{equation}
where $Z_m$ and $Z_{\rm cont}$ are the lattice and continuum
renormalization constants~\cite{Broadhurst}, and $\mu$ is the scale at
which the $\overline{MS}$ mass is defined.   The perturbative series
for both $Z_m$ and $Z_{\rm cont}$ have renormalon ambiguities, therefore
so does $m_{pole}$. However, in the desired relation,
\begin{equation}
m_{\overline{MS}}(\mu) = Z^{-1}_{\rm cont}(\mu) Z_m M_b^0 \,,
\label{MSmass}
\end{equation} 
$Z^{-1}_{\rm cont} Z_m $ is ambiguity free.  

We calculate $m_{pole}$ on the lattice in two ways analogous to a
previous determination using the $\Upsilon$ system~\cite{Mb94NRQCD}. In the
first method, we use Eq.~(\ref{eq:Deltadef}) and write $m_{pole} =
M_{meson} - E_{\rm sim} + E_0$ where $M_{meson}$ is the experimental mass,
$E_{\rm sim}$ is measured from the 2-point correlators, and $E_0$ is
calculated using perturbation theory.
The second method, $m_{pole} = Z_m M_b^0$,  uses the perturbative expression for
$Z_m$. The quantities $Z_m$, and $E_0$, calculated to $O(\alpha_s)$,
are listed in Table~\ref{tab:pert} for the different values of $aM^0$.  
The results for the two ways of fixing $M_b^0$ are given in Table~\ref{tab:bquark}. 

This pole mass is converted, as in~\cite{kent}, to
$m_{\overline{MS}}(\mu) = Z^{-1}_{\rm cont}(\mu) m_{pole}$ using
continuum perturbation theory for $Z^{-1}_{\rm cont}(\mu)$ and the
Brodsky-Lepage-Mackenzie procedure~\cite{BLM} to set the scale for the
coupling constant.  For $\mu$ we choose values between $1/a$ and
$\pi/a$, avoiding those values where the BLM procedure fails. We then
use 2-loop running to get the final result
$m_{\overline{MS}}(m_{\overline{MS}})$, which, in principle, should
not depend on the choice of the intermediate scale $\mu$. These
results are also given in Table~\ref{tab:bquark}, where the second
error is the spread with respect to varying $\mu$, and is indicative
of the neglect of the higher order terms in the perturbative
expressions.

Our preferred 
determination of $m_{\overline{MS}}(\mu)$ comes from ``directly'' expanding
the product $Z^{-1}_{\rm cont}(\mu) Z_m$ in Eq.~(\ref{MSmass}) to
$O(\alpha_s)$~\cite{kent} and using the Lepage-Mackenzie 
procedure~\cite{lepmac} to 
calculate the appropriate scale $q^*$ at which to evaluate
$\alpha_s$~\cite{kent}. The reason for choosing this as the preferred method,
as explained before, is the cancellation of renormalons in the product
and the much better value of $q^*$. 
Continuum ($\overline{MS}$) running is then used to convert
$m_{\overline{MS}}(\mu)$ to $m_{\overline{MS}}(m_{\overline{MS}})$.  
Our final result, obtained by fixing $M^0_b$ from the spin-averaged
$M^\prime$, is 
\begin{equation}
m_{\overline{MS}} (m_{\overline{MS}}) = 4.35(10)({}^{-3}_{+2})(20) \GeV,
\end{equation}
where the first error includes statistics and interpolation
uncertainty; the second is from the uncertainty in the lattice spacing; 
and the third is the systematic error associated with using one-loop
perturbation theory. We estimate it as being $1 \times \alpha_s^2$.
For typical values of $\alpha_s$, depending on the matching scale $\mu$,
this is $\sim 2.5-5\%$. To be conservative, we assign a 200 MeV
perturbative error to the mass.

There are two previous lattice determinations of $m_b$ using a one-loop
matching procedure. The NRQCD
collaboration~\cite{Mb94NRQCD,kent,PDB98} has calculated it within the
$\Upsilon$ system,
and the APE collaboration~\cite{Mb97HQET,PDB98} evaluates $E_{\rm sim}
- E_0$ for the $B$ meson in the static theory. In addition, the APE group has
recently extended their matching calculation to two
loops~\cite{APE98}.  These three results are
\begin{eqnarray}
  m_b^{\MSbar}(m_b^{\MSbar}) &=& 4.16(5)(15)\ \GeV    \quad{\rm (NRQCD, 1-loop) }, \nonumber\\
                             &=& 4.15(5)(20)\ \GeV \quad{\rm (APE, 1-loop)}, \nonumber 
\\
                             &=& 4.41(5)(10)\ \GeV \quad{\rm (APE, 2-loop)}.
\end{eqnarray}
While all these results are consistent within errors, a couple of points are
in order.  First, the results of the APE calculation, which is similar to our method 1, 
suggest that the 2-loop term is large.
This is consistent with our finding that the $a q^\ast$ for $E_0$ (and
$Z_m$) is small, $\sim 0.6$. Such a small value of $a q^\ast$ is
indicative of a large coefficient of the 2-loop term in the bubble
summation approximation (BSA). Thus in Methods 1 and 2, our estimate of
perturbative uncertainty in the mass, due to the large value of $ 1
\times \alpha_s^2(q^\ast)$, is $\sim 400 $ MeV.  In our preferred direct
method,  $Z^{-1}_{\rm cont}(\mu) Z_m$ has no renormalons, and the series is
better behaved in the BSA. Our estimate of the uncertainty, $200$ MeV, 
is based on the correspondingly larger value of $q^\ast$. To go beyond such an
order of magnitude estimate, a two-loop
calculation needs to be done within NRQCD since the 1-loop calculation shows a 
strong dependence of the coefficient on $aM^0$. 
Second, we find that the variation of $E_{\rm bind} \equiv E_{\rm sim} - E_0$ 
with $a M^0$ is small, i.e. $O(50)$ MeV (see Table~\ref{tab:LambdabarB}).
We estimate that the $O(\Lambda_{QCD}^2/M)$ corrections to the APE results
are of this order. Thus, we expect the systematic error in the APE
calculation~\cite{APE98} to be slightly smaller than ours. We shall 
present a more detailed comparison of $m_{\overline{MS}}$ 
from  the heavy-light and heavy-heavy systems on the same configurations
in a separate publication~\cite{Mbfuture}.

\begin{table}[thbp]
\begin{center}
\begin{tabular}{|c|c|c|c|c|c|}
\noalign{\hrule}
\multicolumn{1}{|c|}{} &
\multicolumn{2}{c|}{} &
\multicolumn{3}{c|}{} \\[-7pt]
\multicolumn{1}{|c|}{Method to} &
\multicolumn{2}{c|}{Pole mass [GeV]} &
\multicolumn{3}{|c|}{$\overline{MS}$ mass [GeV]} \\
\multicolumn{1}{|c|}{fix $aM^0_b$} &
\multicolumn{1}{c|}{Method 1} &
\multicolumn{1}{c|}{Method 2} &
\multicolumn{1}{c|}{Method 1} &
\multicolumn{1}{c|}{Method 2} &
\multicolumn{1}{c|}{Direct} \\
\noalign{\hrule}
$M^\prime$ (spin-avg $B$)  & 4.96(1)  & 4.97(10)   & 4.43(1)(4)  &  4.44(10)(4)  & 4.35(10)(4)   \\
\noalign{\hrule}                                             
$M_{\rm kin}$ ($B$ )       & 4.96(3)  & 4.76(41)   & 4.43(2)(4)  &  4.25(37)(4)  & 4.15(38)(4)   \\
\noalign{\hrule}                                             
\end{tabular}
\end{center}
\caption{Results for the $b$ quark pole and $\overline{MS}$
masses. Method 1 uses the meson mass and $E_0$, while method 2 uses
$Z_m$ and $M_b^0$. Both methods are described in more detail
in~\protect\cite{Mb94NRQCD}. The Direct method is described in the
text. The first error quoted is statistical and
includes interpolation/extrapolation to the physical quark masses; 
the second is due to the variation in the matching scale $\mu$.
}
\label{tab:bquark}
\end{table}

\section{Heavy-light mesons}
\label{sec:mesons}

The bare lattice results for meson energies and splittings as a
function of $\kappa$ and $aM^0$ are presented in
Tables~\ref{tab:sstates} and~\ref{tab:pstates}. These are first
extrapolated/interpolated linearly to $\kappa_l$ and $\kappa_s$, and
then to $aM_b^0$ to obtain estimates for the physical states. (The
data are not precise enough to include higher order corrections in the
fits.) To show the dependence of the mass splittings on the heavy
quark mass we plot them as a function of $1/\overline{M} \equiv
4/(3M_{H^\ast} + M_H)$. In this paper,  we use $h$ to denote a generic 
heavy quark, $H$ for a 
heavy-light meson,  and an overbar for spin-averaged quantities. 
Where we find a significant $\overline M$
dependence, we quote the intercept (value in the static limit) and the
slope. In cases where we find no significant slope, we do not show
the corresponding fits in the figures. In general we find that the slope
is $\sim \Lambda_{QCD}^2$, $i.e.$ the corrections to the static
limit are $\sim 10\%$ at $M_b$.

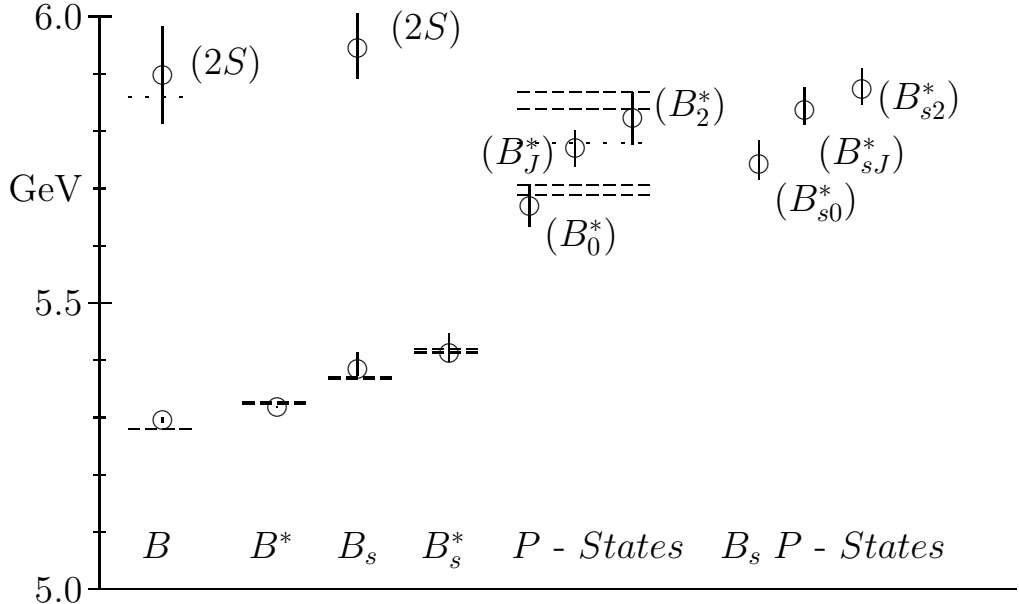
\begin{figure}
\begin{center}
\setlength{\unitlength}{.03in}
\begin{picture}(130,100)(30,500)

\put(15,500){\line(0,1){100}}
\multiput(13,500)(0,50){3}{\line(1,0){4}}
\multiput(14,500)(0,10){11}{\line(1,0){2}}
\put(12,500){\makebox(0,0)[r]{{\large5.0}}}
\put(12,550){\makebox(0,0)[r]{{\large5.5}}}
\put(12,600){\makebox(0,0)[r]{{\large 6.0}}}
\put(12,570){\makebox(0,0)[r]{{\large GeV}}}
\put(15,500){\line(1,0){160}}


     \put(25,510){\makebox(0,0)[t]{{\large $B$}}}
     \put(26,529.5){\circle{3}}
     \put(26,529.5){\line(0,1){0.4}}
     \put(26,529.5){\line(0,-1){0.4}}
     \multiput(20,527.9)(3,0){4}{\line(1,0){2}}

     \put(26,589.8){\circle{3}}
     \put(26,589.8){\line(0,1){8.5}}
     \put(26,589.8){\line(0,-1){8.5}}
     \put(37,595){\makebox(0,0)[t]{{\large $(2S)$}}}
     \multiput(20,586)(3,0){4}{\line(1,0){0.5}}

     \put(45,510){\makebox(0,0)[t]{{\large $B^{*}$}}}
     \put(46,531.9){\circle{3}}
     \put(46,531.9){\line(0,1){0.1}}
     \put(46,531.9){\line(0,-1){0.1}}
     \multiput(40,532.6)(3,0){4}{\line(1,0){2}}
     \multiput(40,532.4)(3,0){4}{\line(1,0){2}}

     \put(60,510){\makebox(0,0)[t]{{\large $B_s$}}}
     \put(60,538.4){\circle{3}}
     \put(60,538.4){\line(0,1){3.0}}
     \put(60,538.4){\line(0,-1){1.1}}
     \multiput(55,537.1)(3,0){4}{\line(1,0){2}}
     \multiput(55,536.7)(3,0){4}{\line(1,0){2}}

     \put(60,594.5){\circle{3}}
     \put(60,594.5){\line(0,1){6.1}}
     \put(60,594.5){\line(0,-1){5.3}}
     \put(72,601){\makebox(0,0)[t]{{\large $(2S)$}}}

     \put(75,510){\makebox(0,0)[t]{{\large $B^{*}_s$}}}
     \put(76,541.2){\circle{3}}
     \put(76,541.2){\line(0,1){3.4}}
     \put(76,541.2){\line(0,-1){1.4}}
     \multiput(70,541.9)(3,0){4}{\line(1,0){2}}
     \multiput(70,541.3)(3,0){4}{\line(1,0){2}}

     \put(102,510){\makebox(0,0)[t]{{\large $P$ - $States$}}}
     \put(90,567.0){\circle{3}}
     \put(90,567.0){\line(0,1){3.7}}
     \put(90,567.0){\line(0,-1){3.7}}
     \put(99,565.0){\makebox(0,0)[t]{{\large $(B^\ast_0)$}}}
     \put(98,577.0){\circle{3}}
     \put(98,577.0){\line(0,1){3.1}}
     \put(98,577.0){\line(0,-1){3.1}}
     \put(88,576){\makebox(0,0){{\large $(B_J^\ast)$}}}
     \put(108,582.2){\circle{3}}
     \put(108,582.2){\line(0,1){4.5}}
     \put(108,582.2){\line(0,-1){4.5}}
     \put(118,584.2){\makebox(0,0){{\large $(B^*_2)$}}}
     \multiput(88,568.8)(3,0){8}{\line(1,0){2}}
     \multiput(88,570.6)(3,0){8}{\line(1,0){2}}
     \multiput(88,577.9)(3,0){8}{\line(1,0){0.5}}

     \put(143,510){\makebox(0,0)[t]{{\large $B_s\;P$ - $States$}}}
     \put(130,574.2){\circle{3}}
     \put(130,574.2){\line(0,1){4.2}}
     \put(130,574.2){\line(0,-1){2.7}}
     \put(140,571.0){\makebox(0,0)[t]{{\large $(B^*_{s0})$}}}
     \put(138,583.6){\circle{3}}
     \put(138,583.6){\line(0,1){3.9}}
     \put(138,583.6){\line(0,-1){2.5}}
     \put(148,576){\makebox(0,0){{\large $(B_{sJ}^\ast)$}}}
     \put(148,587.3){\circle{3}}
     \put(148,587.3){\line(0,1){3.7}}
     \put(148,587.3){\line(0,-1){2.6}}
     \put(158,585.5){\makebox(0,0){{\large $(B^*_{s2})$}}}
     \multiput(88,586.8)(3,0){8}{\line(1,0){2}}
     \multiput(88,583.8)(3,0){8}{\line(1,0){2}}

\end{picture}
\end{center}
\caption{Overview of the $B$ meson spectrum. Circles denote lattice
results, dashed lines give the range of experimental
values~\protect\cite{PDB98}, and the dotted lines indicate preliminary
experimental estimates~\protect\cite{lep1}. Errors include statistics and 
the uncertainty in $\kappa_s$. The 
variation of $a^{-1}$ between 1.8 and 2.0 GeV is not included.
}
\label{fig:mesons}
\end{figure}

A summary of our results at the $b$ mass is presented in Table~\ref{tab:meson_summary}
and compared with experimental data in Fig.~\ref{fig:mesons}.  We find
that the radial and orbital splittings are in agreement with the
preliminary experimental results. The hyperfine splittings $M_{B^\ast}
- M_B$ and $M_{B_s^\ast} - M_{B_s}$ are underestimated as will be
discussed below.  We are able to resolve the $P$ state fine structure
for the first time on the lattice; previous lattice calculations were
done in the static limit and found no significant
splittings~\cite{chrism,duncanetal}.  There has been some controversy
about the ordering of these states in potential model
calculations~\cite{potentialinvert}. We find that the $B^{*}_0$ is the
lightest and $B^{*}_2$ is the heaviest.  Details of the analyses follow.

\begin{table}
 \begin{center}
 \setlength\tabcolsep{0.15cm}
 \begin{tabular}{|c|c|c|c|c|c|c|}
 \noalign{\hrule}
 $aM^0$ & $\kappa$&\ensuremath{aE_{\rm sim}[1S(^1S_0)]}&
 \ensuremath{aE_{\rm sim}[2S(^1S_0)]}&
 \ensuremath{a\Delta E(2S - 1S)}&
 \ensuremath{a(M_{H^\ast} -  M_H)} &
 \ensuremath{a\overline{E}} \\
\hline
  1.6& 0.13690&0.493(03)&0.786(23)&0.293(24)&0.020(01)&0.508(03)\\
  2.0&        &0.509(03)&0.795(23)&0.286(24)&0.017(01)&0.522(03)\\
  2.7&        &0.522(03)&0.805(24)&0.283(25)&0.013(01)&0.532(03)\\
  4.0&        &0.531(03)&0.818(24)&0.287(25)&0.009(01)&0.538(03)\\
  7.0&        &0.534(04)&0.805(25)&0.271(26)&0.006(01)&0.538(04)\\
 10.0&        &0.533(04)&0.800(26)&0.267(28)&0.004(01)&0.536(04)\\
\hline                                      
  1.6& 0.13750&0.475(03)&0.770(27)&0.295(28)&0.020(01)&0.490(03)\\
  2.0&        &0.491(03)&0.784(26)&0.293(27)&0.016(01)&0.503(03)\\
  2.7&        &0.505(03)&0.797(29)&0.292(30)&0.013(01)&0.514(03)\\
  4.0&        &0.514(04)&0.798(28)&0.284(30)&0.008(01)&0.520(04)\\
  7.0&        &0.517(05)&0.792(28)&0.275(30)&0.005(01)&0.521(05)\\
 10.0&        &0.517(05)&0.788(30)&0.272(33)&0.004(01)&0.520(05)\\
\hline                                      
  1.6& 0.13808&0.459(04)&0.762(33)&0.303(34)&0.018(02)&0.472(04)\\
  2.0&        &0.476(04)&0.777(35)&0.301(36)&0.015(02)&0.487(05)\\
  2.7&        &0.490(05)&0.788(31)&0.298(32)&0.012(02)&0.499(05)\\
  4.0&        &0.499(05)&0.800(38)&0.302(40)&0.008(01)&0.504(05)\\
  7.0&        &0.501(05)&0.792(34)&0.292(36)&0.006(01)&0.505(05)\\
 10.0&        &0.501(05)&0.786(35)&0.285(37)&0.004(01)&0.504(05)\\
 \noalign{\hrule}
 \end{tabular}
 \end{center}
 \par\vfil\penalty-5000\vfilneg
\caption{$aE_{\rm sim}$ for the $1S$ and $2S$ mesons is obtained using
a two state fit, and $a(M_{H^\ast} - M_{H})$ is obtained from a fit to
the ratio of the correlation functions. The splitting $a\Delta E(2S -
1S)$ and the spin-averaged energy $\overline{E} = [3E_{\rm
sim}({H^\ast}) + E_{\rm sim}(H)]/4$ are calculated within the
bootstrap process.}
\label{tab:sstates}
\end{table}

\begin{table}
 \begin{center}
 \setlength\tabcolsep{0.0cm}
 \begin{tabular}{|c|c|c|c|c|c|c|}
 \noalign{\hrule}
 $aM^0$ & $\kappa$ &\ensuremath{aE_{\rm sim}({}^3P_2T)} &
 \ensuremath{a\Delta E({}^3P_2T-{}^3P_1)}
 &\ensuremath{a\Delta E({}^3P_2T-{}^1P_1)}&\ensuremath{a\Delta
E(^3P_2T-{}^3P_0)} &\ensuremath{a\Delta E(^3P_2T-{}^3P_2E)} \\
\hline
  1.6& 0.13690&0.769(08)&0.042(11)&0.028(07)&0.082(11)&0.020(14)\\
  2.0&        &0.774(06)&0.042(11)&0.028(07)&0.078(11)&0.020(14)\\
  2.7&        &0.772(04)&0.042(11)&0.028(07)&0.073(11)&0.020(13)\\
\hline
  1.6& 0.13750&0.760(09)&0.048(13)&0.032(09)&0.087(12)&0.025(16)\\
  2.0&        &0.765(07)&0.048(13)&0.032(08)&0.083(12)&0.025(16)\\
  2.7&        &0.765(11)&0.048(13)&0.032(08)&0.078(12)&0.025(15)\\
\hline
  1.6& 0.13808&0.752(10)&0.056(16)&0.037(10)&0.093(14)&0.030(20)\\
  2.0&        &0.757(08)&0.055(16)&0.036(10)&0.088(14)&0.029(19)\\
  2.7&        &0.757(12)&0.055(15)&0.036(10)&0.083(14)&0.028(19)\\
 \noalign{\hrule}
 \end{tabular}
 \end{center}
 \par\vfil\penalty-5000\vfilneg
\caption{$aE_{\rm sim}$ and splittings from fits to ratios of
correlators for $P$ states. To obtain the $2^+$ $P$ states we spin-average
over the ${}^3P_2(T)$ and ${}^3P_2(E)$ states.  }
\label{tab:pstates}
\end{table}

\begin{table}[thbp]
\begin{center}
\setlength{\tabcolsep}{0.25cm}
\begin{tabular}{|lc|l|l|}
\hline
\multicolumn{2}{|c|}{state $(n\;J^P)$}&
\multicolumn{1}{c|}{Lattice}&
\multicolumn{1}{c|}{Expt.}\\
\multicolumn{2}{|c|}{}&
\multicolumn{1}{c|}{MeV}&
\multicolumn{1}{c|}{MeV}\\
\hline
\multicolumn{4}{|c|}{heavy-light mesons}\\
\hline
$B    $    & $1({}0^-)$ &  5296(04)($^{-2}_{+3}$)     &    5279      \\
$     $    & $2({}0^-)$ &  5895(116)($^{+20}_{-32}$)  &    5860(*)   \\
$B^*  $    & $1({}1^-)$ &  5319(02)($^{+0}_{-2}$)     &    5325(1)   \\
$B^*_0$    & $1({}0^+)$ &  5670(37)($^{+16}_{-24}$)   &              \\
$B_J^\ast$ &           &  5770(31)($^{+24}_{-35}$)   &    5697(9)  \\
$B^*_2$    & $1({}2^+)$ &  5822(45)($^{+27}_{-35}$)   &    5779(*)\protect\cite{lep1} \\
 &  &  &    5725-5768(*)\protect\cite{ciulli} \\
\hline
\multicolumn{4}{|c|}{heavy-strange mesons}\\
\hline
&&&\\[-12pt]
$B_s  $    & $1({}0^-)$ &  5385(15)($^{-6}_{+7}$)($^{+20}_{-0}$)    &    5369(2)   \\[2pt]
$     $    & $2({}0^-)$ &  5935(57)($^{+27}_{-38}$)($^{+9}_{-0}$)   &              \\[2pt]
$B_s^*$    & $1({}1^-)$ &  5412(14)($^{-4}_{+2}$)($^{+20}_{-0}$)    &    5416(3)   \\[2pt]
$B^*_{s0}$ & $1({}0^+)$ &  5742(27)($^{+14}_{-20}$)($^{+15}_{-0}$)  &              \\[2pt]
$B_{sJ}^\ast$ &   &  5836(25)($^{+20}_{-28}$)($^{+14}_{-0}$)  &    5853(15)  \\[2pt]
$B^*_{s2}$ & $1({}2^+)$ &  5878(26)($^{+23}_{-33}$)($^{+11}_{-0}$)  &              \\[2pt]
\hline
\end{tabular}
\end{center}
\caption{Mass estimates in MeV for various meson states.  The $b$
quark mass is fixed using the spin-averaged $\overline B(1S)$.  The
first error in the lattice data is statistical (including the
statistical error in the lattice spacing), the second comes from
varying $a^{-1}$ between 1.8 and 2.0 GeV, and for the strange mesons,
the third error comes from the uncertainty in the strange quark
mass. Finite lattice volume effects, which could be large for the 
excited states, have not been addressed in this exploratory study. 
Preliminary experimental values are denoted by asterisks.  
The lattice results quoted against the $B_J^\ast$ and $B_{sJ}^\ast$ 
states correspond to  the spin-average of the respective $P$ states, and 
the experimental numbers are for the unresolved broad resonances. 
Unless stated otherwise, experimental
numbers are from the Particle Data Book~\protect\cite{PDB98}.}
\label{tab:meson_summary}
\end{table}

In analyzing the mass splittings, we are motivated by the following
qualitative picture; the mass of a heavy-light hadron is considered 
to be a sum of: 
\begin{itemize}
\item{}
the pole mass of the heavy quark which is 
$\sim 1.5$ GeV for the $c$ quark and $\sim 5.0$ GeV for the $b$; 
\item{}
the constituent mass $m$ of the light quarks which is approximately $300$ MeV
for the $u,d$ and $450$ MeV for the $s$ quark as inferred from the octet and 
decuplet light baryons, and which we expect to give the biggest contribution
to the static binding energy of the ground-state hadrons; 
\item{}
an excitation energy of the light quark, which, for orbitally and 
radially excited states,  we expect  to be of the order of $\Lambda_{QCD}$; 
\item{}
the $O(\Lambda_{QCD}^2/M_h)$ contributions due to the kinetic energy
of the heavy quark and  the heavy-light hyperfine energy   
$E_{\sigma_H \cdot \sigma_l} \approx 45$ MeV, inferred from the  
$B^\ast-B$ splittings; 
\item{}
and a residual binding energy $E_{be}$ encapsulating
the remaining interactions which we expect to be small
[$O(\Lambda_{QCD}^3/M_h^2)$].  
\end{itemize}
We accordingly
construct different linear combinations of meson and baryon masses to
isolate individual terms and estimate their size and dependence on the
quark masses. 

\subsection{\bsbd splitting}
\label{ss:BsBd}


The spin-averaged splitting between $B_s$ and $B_d$ mesons should be
dominated by the difference of the strange and light
quark masses. Our estimate is $ \overline{M}_{B_s} - \overline{M}_{B_d}
 = 90(9)({}^{+5}_{-3})({}^{+20}_{-0})$ MeV, to
be compared to the experimental value $91(3)$ MeV. The largest
uncertainty, the third error, comes from setting $\kappa_s$; estimates using
$\kappa_s(M_{K^\ast}) $ are $\sim 20\%$ higher, a feature seen in all
quenched calculations.

Previous calculations have reported the following results for $M_{B_s} - M_{B_d}$: 
$87({}^{+15}_{-12})({}^{+6}_{-12})$
MeV~\cite{UKQCDmesons}, $86(12)({}^{+7}_{-9})$ MeV~\cite{ESTIAstatic},
and $107(13)$ MeV~\cite{McKELLARnrqcd}. The JLQCD calculation~\cite{JLQCD99}, done at 
$\beta = 5.7$, $5.9$, and $6.1$, sees indications of $\sim 20\%$ scaling violations 
between $\beta = 5.9$ and $6.1$. Averaging the data at the largest two $\beta$ they 
find $87(7)(4)({}^{+19}_{-0})$. For comparison, our result is 
$87(9)({}^{+5}_{-3})({}^{+19}_{-0})$ MeV, and 
the experimental value $90(2)$ MeV~\cite{PDB98}. 

\begin{figure}[thbp]
\begin{center}
\epsfysize=3in
\centerline{\epsfbox{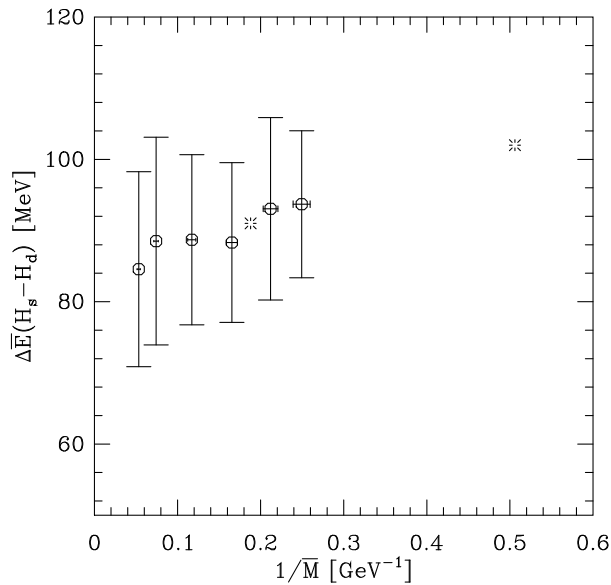 }}
\caption{ Spin-averaged $H_s-H_d$ splitting as a function of the
inverse spin-averaged meson mass $\overline{M}$. The bursts denote the
experimental values for $B$ and $D$ mesons.}
\label{fig:BsbarBdbar}
\end{center}
\end{figure}

In our picture, the heavy quark mass dependence should result from the
difference of the kinetic and hyperfine energies of the heavy quark in
$B_s$ and $B_d$ mesons. (In the spin-averaged splitting, 
$\overline {M}_{B_s} - \overline{M}_{B_d}$, only the difference of the 
kinetic energies remains.) Therefore, we expect this splitting to be
independent of the heavy quark mass up to terms of $O((m_s-m_d)/M)$.   
The experimental
data show a $\sim 10\%$ increase going from the $B$ to the $D$
meson. Our data, given in Table~\ref{tab:BsBd}, show no significant
dependence on the heavy quark mass; however, as shown in
Fig.~\ref{fig:BsbarBdbar}, they are consistent with the experimental
trend. This consistency has also been found in
Ref.~\cite{scalepaper}, where the heavy quark mass dependence has been 
studied at higher statistics and for a heavy quark mass range between the 
$b$ and the $c$.

\begin{table}[thbp]
\lattrue\interfalse 
 \setlength\tabcolsep{0.25cm}
 \begingroup
 \begin{center}
 \lightfalse
 \setupcounts
 \edef\selectcols#1#2#3#4#5#6#7{\selectcols{#1}{#2}{#3}{#4}{#5}{#6}{#7}}
 \edef\putrow#1#2#3#4{\putrow{#1}{#2}{#3}{#4}}
 \begin{tabular}{\putrow{|c|}{r|}%
                 {\selectcols{r|}{r|}{r|}{r|}{r|}{r|}{r|}}%
                 {\selectcols{r|}{r|}{r|}{r|}{r|}{r|}{r|}}}
 \noalign{\hrule}
 \multicolumn{\allcolcount}{|c|}{
 \ensuremath{\Delta E({\overline H}_s - {\overline H}_d)}                    }\\
 \noalign{\hrule}
 \multicolumn{1}{|c|}{} &
 \ifchi\multicolumn{1}{|c|}{} &\fi
 \iflat\multicolumn{\colcount}{c|}{lattice units}&\fi
 \ifMeV\multicolumn{\colcount}{|c|}{MeV}\fi\\
 \noalign{\hrule}
 \putrow{$aM^0$}{&\multicolumn{1}{c|}{\ensuremath{\chi^2}}}%
 {\selectcols{&\multicolumn{1}{c|}{\ensuremath{\kappa_{light}          }}}
 {&\multicolumn{1}{c|}{\ensuremath{\kappa_{av}(m_K)        }}}
 {&\multicolumn{1}{c|}{\ensuremath{\kappa_{av}(m_{K^\ast}) }}}
 {&\multicolumn{1}{c|}{\ensuremath{\kappa_{av}(m_{\phi})   }}}
 {&\multicolumn{1}{c|}{\ensuremath{\kappa_{s}(m_K)         }}}
 {&\multicolumn{1}{c|}{\ensuremath{\kappa_{s}(m_{K^\ast})  }}}
 {&\multicolumn{1}{c|}{\ensuremath{\kappa_{s}(m_\phi)      }}}}%
 {\selectcols{&\multicolumn{1}{c|}{\ensuremath{\kappa_{light}          }}}
 {&\multicolumn{1}{c|}{\ensuremath{\kappa_{av}(m_K)        }}}
 {&\multicolumn{1}{c|}{\ensuremath{\kappa_{av}(m_{K^\ast}) }}}
 {&\multicolumn{1}{c|}{\ensuremath{\kappa_{av}(m_{\phi})   }}}
 {&\multicolumn{1}{c|}{\ensuremath{\kappa_{s}(m_K)         }}}
 {&\multicolumn{1}{c|}{\ensuremath{\kappa_{s}(m_{K^\ast})  }}}
 {&\multicolumn{1}{c|}{\ensuremath{\kappa_{s}(m_\phi)      }}}}\\
 \noalign{\hrule}
 \putrow{1.6                                               }{&$ 0.0    $}%
{\selectcols{&$0.000(00)$}{&$0.022(03)$}{&$0.030(04)$}{&$0.030(04)$}{&$0.049(06)$}{&$0.060(09)$}{&$0.060(09)$}}%
{\selectcols{&$   0(00)$}{&$  43(05)$}{&$  57(07)$}{&$  58(07)$}{&$  94(10)$}{&$ 114(14)$}{&$ 115(14)$}}\\
 \putrow{2.0                                               }{&$ 0.0    $}%
{\selectcols{&$0.000(00)$}{&$0.022(03)$}{&$0.030(05)$}{&$0.030(05)$}{&$0.049(07)$}{&$0.059(10)$}{&$0.060(11)$}}%
{\selectcols{&$   0(00)$}{&$  43(06)$}{&$  57(09)$}{&$  57(09)$}{&$  93(13)$}{&$ 114(17)$}{&$ 114(17)$}}\\
 \putrow{2.7                                               }{&$ 0.0    $}%
{\selectcols{&$0.000(00)$}{&$0.021(03)$}{&$0.028(04)$}{&$0.028(04)$}{&$0.046(06)$}{&$0.056(08)$}{&$0.057(08)$}}%
{\selectcols{&$   0(00)$}{&$  40(05)$}{&$  54(07)$}{&$  54(07)$}{&$  88(11)$}{&$ 108(14)$}{&$ 108(14)$}}\\
 \putrow{4.0                                               }{&$ 0.0    $}%
{\selectcols{&$0.000(00)$}{&$0.021(03)$}{&$0.028(05)$}{&$0.029(05)$}{&$0.046(07)$}{&$0.057(09)$}{&$0.057(09)$}}%
{\selectcols{&$   0(00)$}{&$  41(05)$}{&$  54(08)$}{&$  54(08)$}{&$  89(12)$}{&$ 108(15)$}{&$ 109(15)$}}\\
 \putrow{7.0                                               }{&$ 0.0    $}%
{\selectcols{&$0.000(00)$}{&$0.021(04)$}{&$0.028(05)$}{&$0.028(05)$}{&$0.046(08)$}{&$0.056(11)$}{&$0.057(11)$}}%
{\selectcols{&$   0(00)$}{&$  41(07)$}{&$  54(09)$}{&$  54(09)$}{&$  89(15)$}{&$ 108(18)$}{&$ 109(18)$}}\\
 \putrow{10.0                                              }{&$ 0.0    $}%
{\selectcols{&$0.000(00)$}{&$0.020(03)$}{&$0.027(05)$}{&$0.027(05)$}{&$0.044(07)$}{&$0.054(09)$}{&$0.054(10)$}}%
{\selectcols{&$   0(00)$}{&$  39(06)$}{&$  51(08)$}{&$  52(08)$}{&$  85(14)$}{&$ 103(16)$}{&$ 104(16)$}}\\
 \noalign{\hrule}
 \end{tabular}
 \end{center}
 \endgroup
 \par\vfil\penalty-5000\vfilneg

\caption{Spin-averaged $H_s-H_d$ splitting as a function of
$M^0$. The experimental value is $91(3)$ MeV.}
\label{tab:BsBd}
\end{table}

\subsection{\sss splitting}
\label{ss:2S1S}

\begin{figure}[thbp]
\begin{center}
\epsfysize=3in
\centerline{\epsfbox{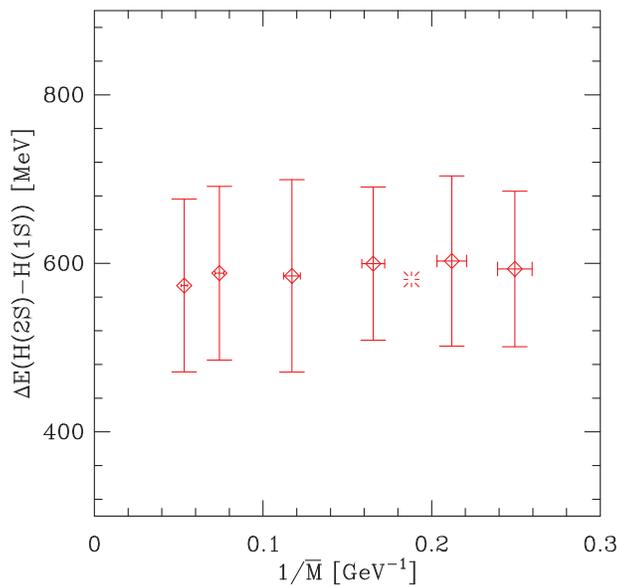 }}
\caption{ $2S-1S$ splitting for $^1S_0$ 
states as a function of the inverse spin-averaged meson
mass. The burst denotes the preliminary $^1S_0$ experimental
value~\protect\cite{lep1}.}
\label{fig:2S1S}
\end{center}
\end{figure}

The raw data for the $2\,{}^1S_0-1\,{}^1S_0$ splitting are given in
Table~\ref{tab:sstates}, and after extrapolation or interpolation to
$\kappa_l$ and $\kappa_s$, in Table~\ref{tab:2S1S}.  This splitting
should be dominated by the difference in the kinetic energies of the
light and the heavy quarks which give contributions of
$O(\Lambda_{QCD}^2/m_{constituent})$ and $O(\Lambda_{QCD}^2/M)$
respectively. With our data, as shown in Table~\ref{tab:2S1S} and
illustrated in Fig.~\ref{fig:2S1S}, we cannot resolve any dependence
on either the light or the heavy quark mass.  Our results for $B$ and
$B_s$ systems are
$602(86)({}^{+25}_{-35})$ and
$559(55)({}^{+31}_{-38})({}^{+0}_{-12})$ MeV respectively, to be
compared with the preliminary experimental value, $581$ MeV, for the
$B$~\cite{lep1}.  In the charm sector, the most relevant experimental value is
$627$ MeV for the $D^{\ast\prime}$~\cite{abreu98}.  

We do not give results for the spin-averaged splitting $2\overline S -
1\overline S$, since the signal for the $^3S_1$ excited state is less 
reliable than that for the $^1S_0$.


\begin{table}[ht]
\lattrue\interfalse 
 \setlength\tabcolsep{0.25cm}
 \begingroup
 \begin{center}
 \setupcounts
 \edef\selectcols#1#2#3#4#5#6#7{\selectcols{#1}{#2}{#3}{#4}{#5}{#6}{#7}}
 \edef\putrow#1#2#3#4{\putrow{#1}{#2}{#3}{#4}}
 \begin{tabular}{\putrow{|c|}{r|}%
                 {\selectcols{r|}{r|}{r|}{r|}{r|}{r|}{r|}}%
                 {\selectcols{r|}{r|}{r|}{r|}{r|}{r|}{r|}}}
 \noalign{\hrule}
 \multicolumn{\allcolcount}{|c|}{
 \ensuremath{\Delta E(2\;{}^1S_0 - 1\;{}^1S_0)}    }\\
 \noalign{\hrule}
 \multicolumn{1}{|c|}{} &
 \ifchi\multicolumn{1}{|c|}{} &\fi
 \iflat\multicolumn{\colcount}{c|}{lattice units}&\fi
 \ifMeV\multicolumn{\colcount}{|c|}{MeV}\fi\\
 \noalign{\hrule}
 \putrow{$a M^0$}{&\multicolumn{1}{c|}{\ensuremath{\chi^2}}}%
 {\selectcols{&\multicolumn{1}{c|}{\ensuremath{\kappa_{light}          }}}
 {&\multicolumn{1}{c|}{\ensuremath{\kappa_{av}(m_K)        }}}
 {&\multicolumn{1}{c|}{\ensuremath{\kappa_{av}(m_{K^\ast}) }}}
 {&\multicolumn{1}{c|}{\ensuremath{\kappa_{av}(m_{\phi})   }}}
 {&\multicolumn{1}{c|}{\ensuremath{\kappa_{s}(m_K)         }}}
 {&\multicolumn{1}{c|}{\ensuremath{\kappa_{s}(m_{K^\ast})  }}}
 {&\multicolumn{1}{c|}{\ensuremath{\kappa_{s}(m_\phi)      }}}}%
 {\selectcols{&\multicolumn{1}{c|}{\ensuremath{\kappa_{light}          }}}
 {&\multicolumn{1}{c|}{\ensuremath{\kappa_{av}(m_K)        }}}
 {&\multicolumn{1}{c|}{\ensuremath{\kappa_{av}(m_{K^\ast}) }}}
 {&\multicolumn{1}{c|}{\ensuremath{\kappa_{av}(m_{\phi})   }}}
 {&\multicolumn{1}{c|}{\ensuremath{\kappa_{s}(m_K)         }}}
 {&\multicolumn{1}{c|}{\ensuremath{\kappa_{s}(m_{K^\ast})  }}}
 {&\multicolumn{1}{c|}{\ensuremath{\kappa_{s}(m_\phi)      }}}}\\
 \noalign{\hrule}
 \putrow{1.6                                               }{&$0.13    $}%
{\selectcols{&$0.310(49)$}{&$0.304(38)$}{&$0.302(34)$}{&$0.302(34)$}{&$0.297(27)$}{&$0.294(24)$}{&$0.294(24)$}}%
{\selectcols{&$ 593( 92)$}{&$ 583( 72)$}{&$ 579( 66)$}{&$ 579( 66)$}{&$ 570( 53)$}{&$ 564( 49)$}{&$ 564( 48)$}}\\
 \putrow{2.0                                               }{&$0.11    $}%
{\selectcols{&$0.315(54)$}{&$0.305(40)$}{&$0.302(36)$}{&$0.302(35)$}{&$0.294(27)$}{&$0.289(24)$}{&$0.288(24)$}}%
{\selectcols{&$ 603(101)$}{&$ 585( 76)$}{&$ 578( 69)$}{&$ 578( 68)$}{&$ 563( 54)$}{&$ 553( 50)$}{&$ 553( 50)$}}\\
 \putrow{2.7                                               }{&$0.17    $}%
{\selectcols{&$0.313(46)$}{&$0.303(36)$}{&$0.300(32)$}{&$0.300(32)$}{&$0.292(27)$}{&$0.287(25)$}{&$0.287(25)$}}%
{\selectcols{&$ 600( 91)$}{&$ 581( 71)$}{&$ 575( 66)$}{&$ 575( 66)$}{&$ 560( 55)$}{&$ 551( 53)$}{&$ 551( 53)$}}\\
 \putrow{4.0                                               }{&$0.25    $}%
{\selectcols{&$0.305(59)$}{&$0.299(44)$}{&$0.296(40)$}{&$0.296(40)$}{&$0.291(30)$}{&$0.287(26)$}{&$0.287(26)$}}%
{\selectcols{&$ 585(114)$}{&$ 573( 86)$}{&$ 568( 78)$}{&$ 568( 78)$}{&$ 558( 60)$}{&$ 551( 54)$}{&$ 551( 53)$}}\\
 \putrow{7.0                                               }{&$0.24    $}%
{\selectcols{&$0.307(53)$}{&$0.295(40)$}{&$0.290(37)$}{&$0.290(37)$}{&$0.280(29)$}{&$0.274(27)$}{&$0.273(27)$}}%
{\selectcols{&$ 588(103)$}{&$ 565( 80)$}{&$ 556( 74)$}{&$ 556( 74)$}{&$ 537( 60)$}{&$ 525( 57)$}{&$ 524( 57)$}}\\
 \putrow{10.0                                              }{&$0.22    $}%
{\selectcols{&$0.299(53)$}{&$0.288(41)$}{&$0.284(38)$}{&$0.284(38)$}{&$0.275(31)$}{&$0.270(29)$}{&$0.269(29)$}}%
{\selectcols{&$ 574(103)$}{&$ 552( 82)$}{&$ 545( 76)$}{&$ 545( 76)$}{&$ 527( 64)$}{&$ 517( 62)$}{&$ 517( 62)$}}\\
 \noalign{\hrule}
 \end{tabular}
 \end{center}
 \endgroup
 \par\vfil\penalty-5000\vfilneg

\caption{$2S-1S$ splittings extrapolated/interpolated to $\kappa_l$
and $\kappa_s$.  The preliminary experimental result for $B_d$ is 581
MeV~\protect\cite{lep1}, while for $B_s$ there is no result as yet.}
\label{tab:2S1S}
\end{table}


\subsection{\pminuss splitting}
\label{ss:1P1S}

The two main contributions to the spin-averaged $1P-1S$ splitting
should be the energy it takes to excite the light quark to angular
momentum one, $O(\Lambda_{QCD})$, and the difference of the kinetic
energy of the heavy quark in an $S$-wave and a $P$-wave light quark
background, $O(\Lambda_{QCD}^2/M)$.  Our results, shown in
Table~\ref{tab:PbarSbar}, are constructed from the raw data given in
Tables~\ref{tab:sstates} and \ref{tab:pstates}. Our estimates 
are $457(31)({}^{+24}_{-35})$ MeV for the $B$, and
$428(27)({}^{+27}_{-41})({}^{+0}_{-2})$ MeV for the $B_s$.

Experimentally the $P$ states have not been resolved.  The $P$ wave
resonances $B_J^\ast(5732)$ (or $B^{\ast\ast}$) at $5697(9)$ MeV and $B_{sJ}^\ast(5850)$
at $5853(15)$ MeV are expected to be a superposition of the various
$P$ states. These are $419$ and $484$ MeV higher than the
corresponding ${}^1S_0$ states.  We use them as estimates of the
spin-averaged $1P-1S$ splittings to compare against.

\begin{figure}[thbp]
\begin{center}
\epsfysize=3in
\centerline{\epsfbox{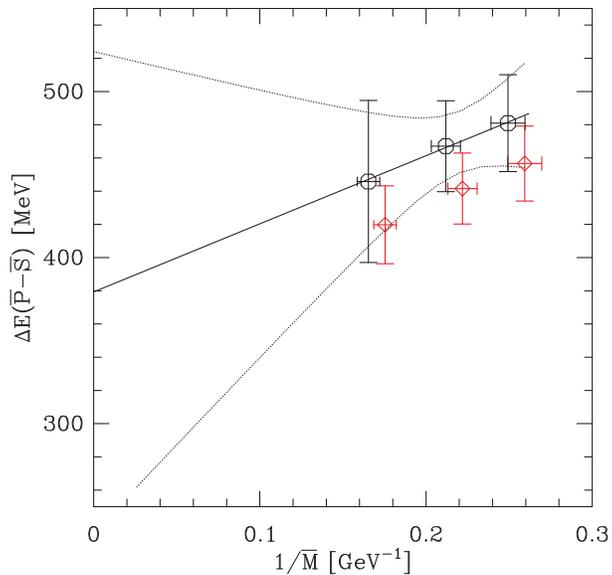 }}
\caption{ Spin-averaged $P-S$ splitting of the $H_d$ (circles) and
$H_s$ mesons (diamonds)
as a function of the inverse spin-averaged meson mass. The lines denote a 
linear fit to the $H_d$ data.}
\label{fig:PbarSbar}
\end{center}
\end{figure}

The variation with either the heavy or the light quark mass is similar
to that in the $2S-1S$ splitting. There is a small decrease with increasing 
light quark mass.  The slope, as a function of
$1/\overline{M}$, is $0.380(202)({}^{+53}_{-66})({}^{+68}_{-0})$
GeV$^2$ for $\kappa_l$, and an almost identical value at $\kappa_s$, as shown in
Fig.~\ref{fig:PbarSbar}.


\begin{table}[ht]
\lattrue\interfalse 
 \setlength\tabcolsep{0.25cm}
 \begingroup
 \begin{center}
 \setupcounts
 \edef\selectcols#1#2#3#4#5#6#7{\selectcols{#1}{#2}{#3}{#4}{#5}{#6}{#7}}
 \edef\putrow#1#2#3#4{\putrow{#1}{#2}{#3}{#4}}
 \begin{tabular}{\putrow{|c|}{r|}%
                 {\selectcols{r|}{r|}{r|}{r|}{r|}{r|}{r|}}%
                 {\selectcols{r|}{r|}{r|}{r|}{r|}{r|}{r|}}}
 \noalign{\hrule}
 \multicolumn{\allcolcount}{|c|}{
 \ensuremath{\Delta E(\overline{P}-\overline{S})}            }\\
 \noalign{\hrule}
 \multicolumn{1}{|c|}{} &
 \ifchi\multicolumn{1}{|c|}{} &\fi
 \iflat\multicolumn{\colcount}{c|}{lattice units}&\fi
 \ifMeV\multicolumn{\colcount}{|c|}{MeV}\fi\\
 \noalign{\hrule}
 \putrow{$a M^0$}{&\multicolumn{1}{c|}{\ensuremath{\chi^2}}}%
 {\selectcols{&\multicolumn{1}{c|}{\ensuremath{\kappa_{light}          }}}
 {&\multicolumn{1}{c|}{\ensuremath{\kappa_{av}(m_K)        }}}
 {&\multicolumn{1}{c|}{\ensuremath{\kappa_{av}(m_{K^\ast}) }}}
 {&\multicolumn{1}{c|}{\ensuremath{\kappa_{av}(m_{\phi})   }}}
 {&\multicolumn{1}{c|}{\ensuremath{\kappa_{s}(m_K)         }}}
 {&\multicolumn{1}{c|}{\ensuremath{\kappa_{s}(m_{K^\ast})  }}}
 {&\multicolumn{1}{c|}{\ensuremath{\kappa_{s}(m_\phi)      }}}}%
 {\selectcols{&\multicolumn{1}{c|}{\ensuremath{\kappa_{light}          }}}
 {&\multicolumn{1}{c|}{\ensuremath{\kappa_{av}(m_K)        }}}
 {&\multicolumn{1}{c|}{\ensuremath{\kappa_{av}(m_{K^\ast}) }}}
 {&\multicolumn{1}{c|}{\ensuremath{\kappa_{av}(m_{\phi})   }}}
 {&\multicolumn{1}{c|}{\ensuremath{\kappa_{s}(m_K)         }}}
 {&\multicolumn{1}{c|}{\ensuremath{\kappa_{s}(m_{K^\ast})  }}}
 {&\multicolumn{1}{c|}{\ensuremath{\kappa_{s}(m_\phi)      }}}}\\
 \noalign{\hrule}
 \putrow{1.6                                               }{&$0.64E-01$}%
{\selectcols{&$0.251(13)$}{&$0.245(11)$}{&$0.243(10)$}{&$0.243(10)$}{&$0.238(08)$}{&$0.235(08)$}{&$0.235(08)$}}%
{\selectcols{&$ 481(29)$}{&$ 470(26)$}{&$ 466(25)$}{&$ 466(25)$}{&$ 457(23)$}{&$ 451(23)$}{&$ 451(23)$}}\\
 \putrow{2.0                                               }{&$0.81E-01$}%
{\selectcols{&$0.244(13)$}{&$0.238(10)$}{&$0.236(09)$}{&$0.235(09)$}{&$0.230(07)$}{&$0.227(07)$}{&$0.227(07)$}}%
{\selectcols{&$ 467(27)$}{&$ 455(24)$}{&$ 452(24)$}{&$ 451(24)$}{&$ 442(21)$}{&$ 436(23)$}{&$ 436(23)$}}\\
 \putrow{2.7                                               }{&$0.26    $}%
{\selectcols{&$0.232(24)$}{&$0.226(16)$}{&$0.224(14)$}{&$0.224(14)$}{&$0.219(08)$}{&$0.216(07)$}{&$0.216(07)$}}%
{\selectcols{&$ 446(49)$}{&$ 434(36)$}{&$ 430(32)$}{&$ 430(32)$}{&$ 420(23)$}{&$ 414(23)$}{&$ 414(23)$}}\\
 \noalign{\hrule}
 \end{tabular}
 \end{center}
 \endgroup
 \par\vfil\penalty-5000\vfilneg

\caption{Spin-averaged $P-S$ splittings. }
\label{tab:PbarSbar}
\end{table}


\subsection{\hyperf splitting}
\label{ss:BstarB}

\begin{figure}[thbp]
\begin{center}
\epsfysize=3in
\centerline{\epsfbox{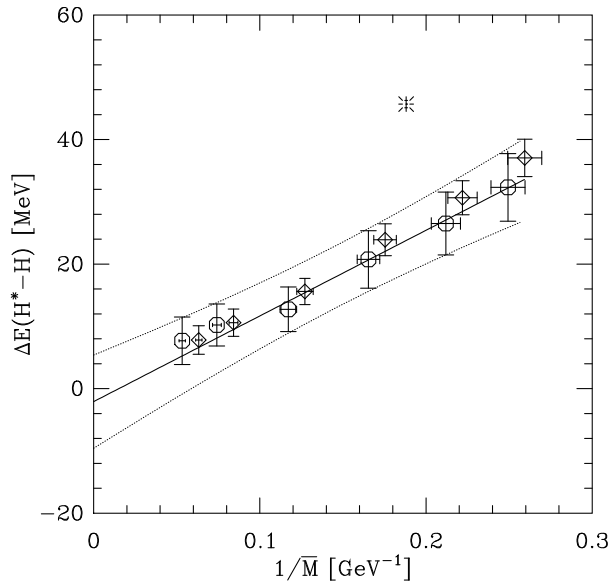 }}
\caption{Hyperfine splitting as a function of the inverse
spin-averaged meson mass.  Circles denote the splitting for $H_d$
mesons, diamonds, for $H_s$ mesons. For clarity, the diamonds are
shifted to the right. The burst denotes the experimental value for
$H_d$ mesons. The lines are a linear fit to the $H_d$ data.}
\label{fig:BstarB}
\end{center}
\end{figure}

Our results for the hyperfine splitting are shown in
Table~\ref{tab:BstarB} and plotted in Fig.~\ref{fig:BstarB}.  A linear
fit to the $B_d$ data gives $0.138(38)({}^{+11}_{-17}) {\rm GeV}^2$
for the slope and $-2(7)$ MeV for the intercept at infinite mass.  A
zero intercept is consistent with the HQET picture in which the
$B^\ast-B$ splitting comes from the interaction of the heavy quark
spin with the color field, $i.e.$ through a $\sigma\cdot B/(2M)$
interaction.  Our
estimates are $24(5)({}^{+2}_{-3})$ MeV and
$27(3)({}^{+2}_{-3})({}^{+1}_{-0})$ MeV for the $B$ and the $B_s$
respectively, and $\Delta E(B^\ast_s-B_s)/\Delta E(B^\ast_d-B_d) =
1.19(20)({}^{-2}_{+2})({}^{+4}_{-0})$.  These splittings are roughly
half the experimental values, $46$ and $47$ MeV respectively.

An underestimate of hyperfine splittings has also been seen by the
previous quenched
calculations~\cite{UKQCDmesons,peterboyle,woloshin,McKELLARnrqcd,JLQCD99}. The 
results of the JLQCD calculation~\cite{JLQCD99} suggest that this is not due 
to scaling violations.  Present preliminary unquenched 
calculations~\cite{sara99} do not show any significant improvement either, 
however, the mass of the two flavors
of dynamical quarks is large, $\sim m_s$. Further work is needed to
clarify this issue.


All hyperfine splittings, including those in the $P$ state and baryon
sector, are, to leading order, generated by the $\sigma\cdot B$ term
in the quark action. It has recently been pointed out that the
coefficient of this term should be larger by a factor of
$1.15-1.30$~\cite{trottier98,sara99}. Such a correction would bring 
the quenched results much closer to the experimental values.



\begin{table}[thbp]
\lattrue\interfalse 
 \setlength\tabcolsep{0.25cm}
 \begingroup
 \begin{center}
 \setupcounts
 \edef\selectcols#1#2#3#4#5#6#7{\selectcols{#1}{#2}{#3}{#4}{#5}{#6}{#7}}
 \edef\putrow#1#2#3#4{\putrow{#1}{#2}{#3}{#4}}
 \begin{tabular}{\putrow{|c|}{r|}%
                 {\selectcols{r|}{r|}{r|}{r|}{r|}{r|}{r|}}%
                 {\selectcols{r|}{r|}{r|}{r|}{r|}{r|}{r|}}}
 \noalign{\hrule}
 \multicolumn{\allcolcount}{|c|}{
 \ensuremath{\Delta E(H^\ast-H)}             }\\
 \noalign{\hrule}
 \multicolumn{1}{|c|}{} &
 \ifchi\multicolumn{1}{|c|}{} &\fi
 \iflat\multicolumn{\colcount}{c|}{lattice units}&\fi
 \ifMeV\multicolumn{\colcount}{|c|}{MeV}\fi\\
 \noalign{\hrule}
 \putrow{$a M^0$}{&\multicolumn{1}{c|}{\ensuremath{\chi^2}}}%
 {\selectcols{&\multicolumn{1}{c|}{\ensuremath{\kappa_{light}          }}}
 {&\multicolumn{1}{c|}{\ensuremath{\kappa_{av}(m_K)        }}}
 {&\multicolumn{1}{c|}{\ensuremath{\kappa_{av}(m_{K^\ast}) }}}
 {&\multicolumn{1}{c|}{\ensuremath{\kappa_{av}(m_{\phi})   }}}
 {&\multicolumn{1}{c|}{\ensuremath{\kappa_{s}(m_K)         }}}
 {&\multicolumn{1}{c|}{\ensuremath{\kappa_{s}(m_{K^\ast})  }}}
 {&\multicolumn{1}{c|}{\ensuremath{\kappa_{s}(m_\phi)      }}}}%
 {\selectcols{&\multicolumn{1}{c|}{\ensuremath{\kappa_{light}          }}}
 {&\multicolumn{1}{c|}{\ensuremath{\kappa_{av}(m_K)        }}}
 {&\multicolumn{1}{c|}{\ensuremath{\kappa_{av}(m_{K^\ast}) }}}
 {&\multicolumn{1}{c|}{\ensuremath{\kappa_{av}(m_{\phi})   }}}
 {&\multicolumn{1}{c|}{\ensuremath{\kappa_{s}(m_K)         }}}
 {&\multicolumn{1}{c|}{\ensuremath{\kappa_{s}(m_{K^\ast})  }}}
 {&\multicolumn{1}{c|}{\ensuremath{\kappa_{s}(m_\phi)      }}}}\\
 \noalign{\hrule}
 \putrow{1.6                                               }{&$0.47E-01$}%
{\selectcols{&$0.017(03)$}{&$0.018(02)$}{&$0.018(02)$}{&$0.018(02)$}{&$0.019(01)$}{&$0.020(01)$}{&$0.020(01)$}}%
{\selectcols{&$  32(05)$}{&$  34(04)$}{&$  35(04)$}{&$  35(04)$}{&$  37(03)$}{&$  38(03)$}{&$  38(02)$}}\\
 \putrow{2.0                                               }{&$0.47E-01$}%
{\selectcols{&$0.014(02)$}{&$0.015(02)$}{&$0.015(02)$}{&$0.015(02)$}{&$0.016(01)$}{&$0.016(01)$}{&$0.017(01)$}}%
{\selectcols{&$  27(05)$}{&$  28(04)$}{&$  29(04)$}{&$  29(04)$}{&$  31(03)$}{&$  32(02)$}{&$  32(02)$}}\\
 \putrow{2.7                                               }{&$0.39E-01$}%
{\selectcols{&$0.011(02)$}{&$0.012(02)$}{&$0.012(02)$}{&$0.012(02)$}{&$0.012(01)$}{&$0.013(01)$}{&$0.013(01)$}}%
{\selectcols{&$  21(05)$}{&$  22(04)$}{&$  23(03)$}{&$  23(03)$}{&$  24(03)$}{&$  25(02)$}{&$  25(02)$}}\\
 \putrow{4.0                                               }{&$0.43E-01$}%
{\selectcols{&$0.007(02)$}{&$0.007(01)$}{&$0.008(01)$}{&$0.008(01)$}{&$0.008(01)$}{&$0.008(01)$}{&$0.009(01)$}}%
{\selectcols{&$  13(04)$}{&$  14(03)$}{&$  15(03)$}{&$  15(03)$}{&$  16(02)$}{&$  16(02)$}{&$  16(02)$}}\\
 \putrow{7.0                                               }{&$0.20E-01$}%
{\selectcols{&$0.005(02)$}{&$0.005(01)$}{&$0.005(01)$}{&$0.005(01)$}{&$0.006(01)$}{&$0.006(01)$}{&$0.006(01)$}}%
{\selectcols{&$  10(03)$}{&$  10(03)$}{&$  10(03)$}{&$  10(03)$}{&$  11(02)$}{&$  11(02)$}{&$  11(02)$}}\\
 \putrow{10.0                                              }{&$0.20E-01$}%
{\selectcols{&$0.004(02)$}{&$0.004(02)$}{&$0.004(01)$}{&$0.004(01)$}{&$0.004(01)$}{&$0.004(01)$}{&$0.004(01)$}}%
{\selectcols{&$   8(04)$}{&$   8(03)$}{&$   8(03)$}{&$   8(03)$}{&$   8(02)$}{&$   8(02)$}{&$   8(02)$}}\\
 \noalign{\hrule}
 \end{tabular}
 \end{center}
 \endgroup
 \par\vfil\penalty-5000\vfilneg

\caption{$H^\ast-H$ splitting as a function of $M^0$. The experimental 
results are $45.78(35)$ MeV for $B_d$ and $47.0(2.6)$ MeV for $B_s$.}
\label{tab:BstarB}
\end{table}

\subsection{$P$ fine structure}
\label{ss:Pfine}

In the $jj$ coupling scheme there are two doublets of $P$ states which
are distinguished by the angular momentum of the light quark: $j_l =
1/2$ and $j_l = 3/2$.  The states in each doublet are separated by a
spin flip of the heavy quark into a $0^+$ and a $1^+$ state for $j_l = 1/2$
($B_0^\ast$ and $B_1^\ast$),
and a $1^{+\prime}$ and a $2^+$ state for $j_l = 3/2$ ($B_1$ and $B_2^\ast$). 
We therefore expect the 
spin-averages of the $j_l = 3/2$ and the $j_l = 1/2$ doublets to be separated
by $O(\Lambda_{QCD})$, and the states within each doublet by 
$O(\Lambda_{QCD}^2/M)$.

The experimental situation is as follows. There exists a broad
resonance at 5697(9) MeV~\cite{PDB98}, whose spin has not been
determined and which is believed to be a superposition of various
$P$ states. There is also a preliminary experimental result by the
DELPHI collaboration~\cite{lep1} for a narrow $P$ state which is 81
MeV heavier than this resonance.  Its spin is also not resolved, but
it is believed to be either $J=1$ or $J=2$.

Recently, estimates for individual $P$ states have been obtained by
fitting the line shape of the broad resonance using phenomenological
input based on HQET for the mass splittings, decay widths, relative
production rates, and branching fractions~\cite{ciulli}.  Using this
method, the CDF and ALEPH collaborations obtain a mass of the
$B_2^\ast$ of $\sim 5730$ MeV. This result seems to be rather
insensitive to the assumption about the $B_2^* - B_1^*$ splitting,
which the phenomenological model predicts to be $\sim 100$ MeV. The L3
collaboration also uses hyperfine splittings of 12 MeV as input, but
makes no assumption about the splitting between the $j_l = 3/2$ and
the $j_l = 1/2$ doublet, and obtains  slightly higher masses, 
$B_2^\ast \sim 5768$ and  $B_1^\ast \sim 5670$ MeV. Our resolution of the $P$
state fine structure is as follows.

\begin{figure}[thbp]
\begin{center}
\epsfysize=3in
\centerline{\epsfbox{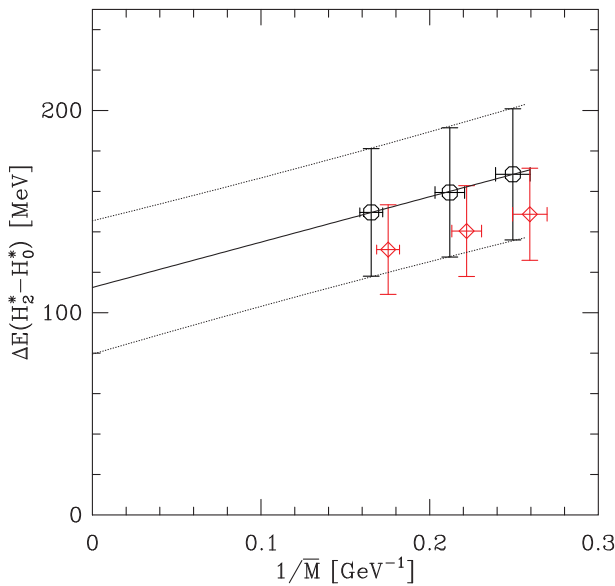 }}
\caption{$H_2^\ast-H^\ast_0$ (denoted by octagons) and $H^*_{s2} -
H^*_{s0}$ (the diamonds are shifted by $0.01$ in the $x$
direction for clarity)
splittings as a function of the inverse spin-averaged meson mass.  A
linear fit to $H_2^\ast-H^\ast_0$ is also shown.}
\label{fig:B2starB0star}
\vspace{-1cm}
\end{center}
\end{figure}

First, we discuss the $0^+$ and $2^+$ states for which the data are
shown in Table~\ref{tab:B2starB0star} and Fig.~\ref{fig:B2starB0star}.
We find that $B_2^\ast-B_0^\ast =
155(32)({}^{+9}_{-13})$ MeV and $B_{s2}^\ast-B_{s0}^\ast
= 136(23)({}^{+10}_{-13})({}^{+0}_{-4})$ MeV.  At $\kappa_l$, the
slope versus $1/\overline{M}$ is $0.224(70)({}^{+20}_{-27}) {\rm
GeV}^2$ and the intercept
is $112(33)({}^{+5}_{-6})$ MeV. For $B_s$ $P$ states, the slope is
$0.209(45)({}^{+19}_{-26})({}^{+0}_{-4}) {\rm GeV}^2$ and the
intercept is $97(23)({}^{+5}_{-8})({}^{+0}_{-4})$ MeV. These results
are a significant improvement over previous values obtained in the
static approach, $i.e.$ $\sim 50(100)$ MeV~\cite{duncanetal} and $\sim
80(75)$ MeV~\cite{chrism} for the intercept.

The situation in model calculations is very unclear. The predictions
are model dependent, and details like the treatment and the 
mass of the light quark are
significant~\cite{potentialinvert}. At this point there is no
consensus on even the sign of the splitting.

\begin{table}[thbp]
\lattrue\interfalse 
 \setlength\tabcolsep{0.25cm}
 \begingroup
 \begin{center}
 \setupcounts
 \edef\selectcols#1#2#3#4#5#6#7{\selectcols{#1}{#2}{#3}{#4}{#5}{#6}{#7}}
 \edef\putrow#1#2#3#4{\putrow{#1}{#2}{#3}{#4}}
 \begin{tabular}{\putrow{|c|}{r|}%
                 {\selectcols{r|}{r|}{r|}{r|}{r|}{r|}{r|}}%
                 {\selectcols{r|}{r|}{r|}{r|}{r|}{r|}{r|}}}
 \noalign{\hrule}
 \multicolumn{\allcolcount}{|c|}{
 \ensuremath{\Delta E(H^\ast_2-H^\ast_0)}            }\\
 \noalign{\hrule}
 \multicolumn{1}{|c|}{} &
 \ifchi\multicolumn{1}{|c|}{} &\fi
 \iflat\multicolumn{\colcount}{c|}{lattice units}&\fi
 \ifMeV\multicolumn{\colcount}{|c|}{MeV}\fi\\
 \noalign{\hrule}
 \putrow{$aM^0$}{&\multicolumn{1}{c|}{\ensuremath{\chi^2}}}%
 {\selectcols
 {&\multicolumn{1}{c|}{\ensuremath{\kappa_{light}          }}}%
 {&\multicolumn{1}{c|}{\ensuremath{\kappa_{av}(m_K)        }}}%
 {&\multicolumn{1}{c|}{\ensuremath{\kappa_{av}(m_{K^\ast}) }}}%
 {&\multicolumn{1}{c|}{\ensuremath{\kappa_{av}(m_{\phi})   }}}%
 {&\multicolumn{1}{c|}{\ensuremath{\kappa_{s}(m_K)         }}}%
 {&\multicolumn{1}{c|}{\ensuremath{\kappa_{s}(m_{K^\ast})  }}}%
 {&\multicolumn{1}{c|}{\ensuremath{\kappa_{s}(m_\phi)      }}}%
 }%
 {\selectcols
 {&\multicolumn{1}{c|}{\ensuremath{\kappa_{light}          }}}%
 {&\multicolumn{1}{c|}{\ensuremath{\kappa_{av}(m_K)        }}}%
 {&\multicolumn{1}{c|}{\ensuremath{\kappa_{av}(m_{K^\ast}) }}}%
 {&\multicolumn{1}{c|}{\ensuremath{\kappa_{av}(m_{\phi})   }}}%
 {&\multicolumn{1}{c|}{\ensuremath{\kappa_{s}(m_K)         }}}%
 {&\multicolumn{1}{c|}{\ensuremath{\kappa_{s}(m_{K^\ast})  }}}%
 {&\multicolumn{1}{c|}{\ensuremath{\kappa_{s}(m_\phi)      }}}%
 }\\
 \noalign{\hrule}
 \putrow{1.6                                               }{&$0.41E-02$}%
{\selectcols{&$0.088(17)$}{&$0.083(14)$}{&$0.082(13)$}{&$0.082(13)$}{&$0.07
8(11)$}{&$0.075(11)$}{&$0.075(11)$}}%
{\selectcols{&$ 168(32)$}{&$ 159(28)$}{&$ 156(26)$}{&$ 156(26)$}{&$ 149(23)
$}{&$ 144(22)$}{&$ 144(22)$}}\\
 \putrow{2.0                                               }{&$0.38E-02$}%
{\selectcols{&$0.083(17)$}{&$0.079(14)$}{&$0.077(13)$}{&$0.077(13)$}{&$0.07
3(11)$}{&$0.071(11)$}{&$0.071(10)$}}%
{\selectcols{&$ 159(32)$}{&$ 151(27)$}{&$ 148(26)$}{&$ 148(26)$}{&$ 140(22)
$}{&$ 136(21)$}{&$ 136(21)$}}\\
 \putrow{2.7                                               }{&$0.34E-02$}%
{\selectcols{&$0.078(16)$}{&$0.074(14)$}{&$0.072(13)$}{&$0.072(13)$}{&$0.06
8(11)$}{&$0.066(10)$}{&$0.066(10)$}}%
{\selectcols{&$ 150(32)$}{&$ 141(27)$}{&$ 138(26)$}{&$ 138(26)$}{&$ 131(22)
$}{&$ 127(21)$}{&$ 127(21)$}}\\
 \noalign{\hrule}
 \end{tabular}
 \end{center}
 \endgroup
 \par\vfil\penalty-5000\vfilneg

\caption{$H_2^\ast-H_0^\ast$ splittings.}
\label{tab:B2starB0star}
\end{table}

To study the $J = 1$ states we used operators with $^3P_1$ and $^1P_1$
quantum numbers in the $LS$ coupling scheme as defined in
Table~\ref{tab:ops}.  The corresponding correlation functions get
contributions from both the physical states. Therefore, at large
Euclidean times both correlators are dominated by the same lowest
state.  The masses we extract from short Euclidean time ($\Delta E t
\ll 1$) correspond to unmixed states in the $LS$ scheme and are not the
physical masses \cite{LANL96HM}.  To get the latter requires a signal
in the mixed correlators followed by a diagonalization of the $2
\times 2$ matrix. Unfortunately, our data does not show a signal in
the mixed correlators, and therefore we do not have results for 
the physical $J=1$ states. The numbers presented in Table~\ref{tab:meson_summary} 
under $1^+$ are those obtained using the ${}^3P_1$ correlators. Estimates 
obtained from the ${}^1P_1$ correlators are almost identical to the 
center of mass of the ${}^3P$ states.

\section{Heavy-light-light baryons}
\label{sec:HLLbaryons}

\begin{figure}
\begin{center}
\setlength{\unitlength}{.025in}
\begin{picture}(130,150)(30,500)
\put(15,500){\line(0,1){150}}
\multiput(13,500)(0,50){4}{\line(1,0){4}}
\multiput(14,500)(0,10){16}{\line(1,0){2}}
\put(12,500){\makebox(0,0)[r]{{\large5.0}}}
\put(12,550){\makebox(0,0)[r]{{\large5.5}}}
\put(12,600){\makebox(0,0)[r]{{\large 6.0}}}
\put(12,650){\makebox(0,0)[r]{{\large 6.5}}}
\put(12,570){\makebox(0,0)[r]{{\large GeV}}}
\put(15,500){\line(1,0){160}}


     \put(31,510){\makebox(0,0)[t]{{\large $B$}}}
     \put(31,529.6){\circle{3}}
     \put(31,529.6){\line(0,-1){0.4}}
     \put(31,529.6){\line(0,1){0.4}}
     \multiput(25,527.9)(3,0){4}{\line(1,0){2}}

     \put(49,510){\makebox(0,0)[t]{{\large $\Lambda_b$}}}
     \put(49,567.9){\circle{3}}
     \put(49,567.9){\line(0,-1){7.1}}
     \put(49,567.9){\line(0,1){7.1}}
     \multiput(43,563.3)(3,0){4}{\line(1,0){2}}
     \multiput(43,561.5)(3,0){4}{\line(1,0){2}}

     \put(67,510){\makebox(0,0)[t]{{\large $\Xi_b$}}}
     \put(67,578.5){\circle{3}}
     \put(67,578.5){\line(0,-1){5.4}}
     \put(67,578.5){\line(0,1){6.9}}

     \put(85,510){\makebox(0,0)[t]{{\large $\Sigma_b$}}}
     \put(85,588.7){\circle{3}}
     \put(85,588.7){\line(0,-1){4.6}}
     \put(85,588.7){\line(0,1){4.6}}
     \multiput(79,580.5)(3,0){4}{\line(1,0){0.5}}
     \multiput(79,578.9)(3,0){4}{\line(1,0){0.5}}

     \put(102,510){\makebox(0,0)[t]{{\large $\Xi'_b$}}}
     \put(102,596.2){\circle{3}}
     \put(102,596.2){\line(0,-1){4.0}}
     \put(102,596.2){\line(0,1){6.4}}

     \put(120,510){\makebox(0,0)[t]{{\large $\Omega_b$}}}
     \put(120,604.8){\circle{3}}
     \put(120,604.8){\line(0,-1){3.3}}
     \put(120,604.8){\line(0,1){6.7}}

     \put(138,510){\makebox(0,0)[t]{{\large $\Sigma_b^*$}}}
     \put(138,590.9){\circle{3}}
     \put(138,590.9){\line(0,1){4.7}}
     \put(138,590.9){\line(0,-1){4.7}}
     \multiput(132,586.1)(3,0){4}{\line(1,0){0.5}}
     \multiput(132,584.5)(3,0){4}{\line(1,0){0.5}}

     \put(156,510){\makebox(0,0)[t]{{\large $\Xi_b^*$}}}
     \put(156,598.2){\circle{3}}
     \put(156,598.2){\line(0,1){6.4}}
     \put(156,598.2){\line(0,-1){3.9}}

     \put(174,510){\makebox(0,0)[t]{{\large $\Omega_b^*$}}}
     \put(174,606.9){\circle{3}}
     \put(174,606.9){\line(0,1){6.9}}
     \put(174,606.9){\line(0,-1){3.4}}

\end{picture}
\end{center}
\caption{Overview of the $b$ baryon spectrum. Circles denote our lattice
results, dashed lines give experimental error bounds~\cite{PDB98}, 
and dotted lines show preliminary experimental results~\protect\cite{lep2,lep1}.
}
\label{fig:HLLbaryons}
\end{figure}
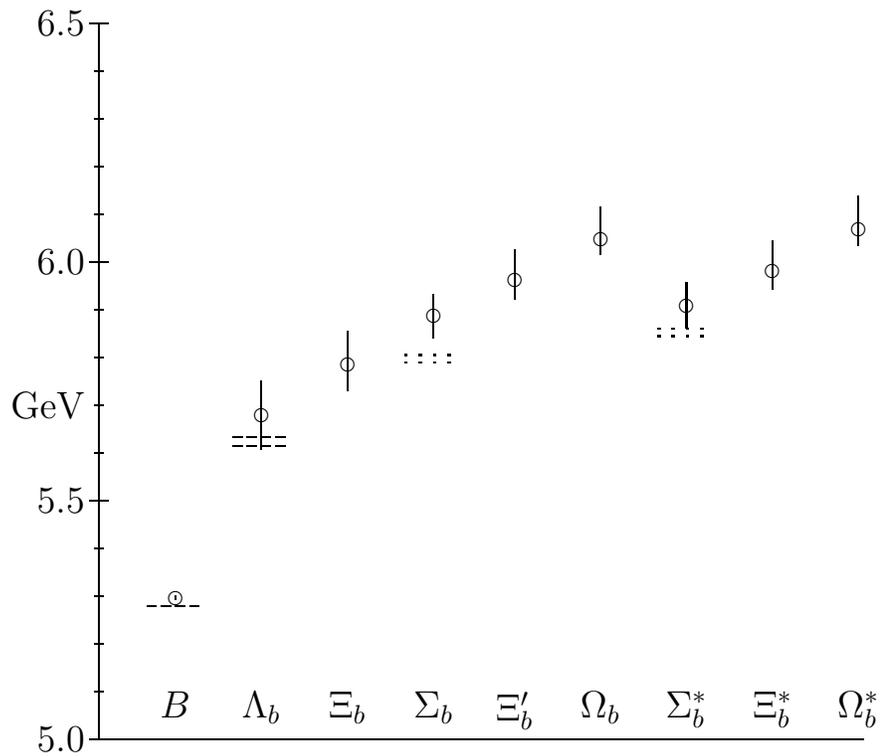

The heavy-light-light baryons, in the heavy quark limit, can be
classified according to the angular momentum of the light quarks. At
zero orbital angular momentum, the light quarks can have total spin
$s_l = 0$ (anti-symmetric in both spin and flavor) and $s_l = 1$
(symmetric in both). As summarized in Table~\ref{tab:baryexp}, there
are three states with $s_l = 0$; $udb$, $usb$, and $dsb$ which are
called the $\Lambda^0_b$, $\Xi^0_b$ and $\Xi^-_b$ baryons with total
spin 1/2. The system with $s_l = 1$ splits up into six hyperfine
doublets, each containing states with spin 1/2 and spin 3/2. These
six doublets are
($\Sigma^+_b, \Sigma^{\ast+}_b$), 
($\Sigma^0_b, \Sigma^{\ast0}_b$), 
($\Sigma^-_b, \Sigma^{\ast-}_b$), 
($\Xi^{\prime0}_b, \Xi^{\ast0}_b$), 
($\Xi^{\prime-}_b, \Xi^{\ast-}_b$), and 
($\Omega^-_b, \Omega^{\ast-}_b$)~\cite{falk97}.  The pairs of states 
($\Sigma^0_b, \Lambda^0_b$), and 
($\Xi^\prime_b, \Xi_b$),  
do not mix if flavor SU(2) is unbroken.  We ensure this in our lattice
calculation by only analyzing baryons with degenerate combinations of
light quarks.  The raw data are given in
Table~\ref{tab:baryons}. Baryons with a generic heavy quark are
denoted as $\Lambda_h$, $\Sigma_h$ etc. To get $us$ and $ds$
combinations we extrapolate linearly in the degenerate light quark
mass to the average mass $(m_s + m_l)/2$, which we label
$\kappa_{av}$. A summary of the experimental numbers and our lattice
results is given in Table~\ref{tab:baryexp} and shown in
Fig.~\ref{fig:HLLbaryons}.

The UKQCD Collaboration has previously presented a similarly detailed
analysis of the baryon spectrum~\cite{UKQCDbaryons}. They used the
tree-level clover action ($C_{SW} = 1$) at $\beta = 6.2$ ($1/a =
2.9(2)$ GeV) and four heavy $\kappa$ around the charm quark mass. In
contrast to our calculation, their $b$ spectrum was obtained by
extrapolation in $1/M$.  To facilitate comparison, we summarize
their results in Table~\ref{tab:baryexp}. Within errors these are
consistent with our findings, although our results are slightly higher
and have a slightly smaller light quark mass dependence.
An important point, as discussed below, is that
we are able to resolve hyperfine splittings for the first time.

The baryon splittings are also analyzed using the phenomenological model discussed
in Sec.~\ref{sec:mesons}. In heavy-light-light baryons there is an
additional light-light hyperfine interaction ($E_{\sigma_l \cdot
\sigma_l}$), which is expected to be of order $\Lambda_{QCD}$.

\begin{table}[tbp]
 \begin{center}
 \begin{tabular}{|c|c|c|c|c|c|}
 \noalign{\hrule}
 $a M^0$ & $\kappa$&\ensuremath{aE_{\rm sim}(\Lambda_h)}&\ensuremath{aE_{\rm sim}(\Sigma_h)}
 &\ensuremath{aE_{\rm sim}(\Sigma^\ast_h)}&
 \ensuremath{a\Delta E(\Sigma^\ast_h-\Sigma_h)} \\
\hline
  1.6& 0.13690&0.801(10)&0.852(10)&0.869(10)&0.014(02)\\
  2.0&        &0.813(11)&0.869(11)&0.883(10)&0.011(02)\\
  2.7&        &0.823(14)&0.884(12)&0.894(10)&0.008(02)\\
  4.0&        &0.823(20)&0.899(14)&0.907(14)&0.005(01)\\
  7.0&        &0.816(29)&0.916(20)&0.921(21)&0.004(01)\\
 10.0&        &0.806(40)&0.929(30)&0.931(32)&0.003(01)\\
\hline
  1.6& 0.13750&0.756(12)&0.822(12)&0.838(12)&0.014(02)\\
  2.0&        &0.769(14)&0.838(13)&0.851(13)&0.011(02)\\
  2.7&        &0.781(17)&0.852(14)&0.860(11)&0.008(02)\\
  4.0&        &0.787(27)&0.866(16)&0.875(19)&0.005(02)\\
  7.0&        &0.782(41)&0.888(23)&0.895(25)&0.004(01)\\
 10.0&        &0.776(56)&0.907(35)&0.912(38)&0.003(01)\\
\hline
  1.6& 0.13808&0.710(17)&0.795(13)&0.812(15)&0.015(03)\\
  2.0&        &0.726(20)&0.811(14)&0.825(15)&0.011(02)\\
  2.7&        &0.739(25)&0.826(17)&0.829(13)&0.008(02)\\
  4.0&        &0.755(41)&0.835(22)&0.843(19)&0.005(02)\\
  7.0&        &0.758(32)&0.865(29)&0.876(34)&0.004(02)\\
 10.0&        &0.766(43)&0.894(43)&0.900(45)&0.002(02)\\
 \noalign{\hrule}
 \end{tabular}
 \end{center}
 \par\vfil\penalty-5000\vfilneg
\caption{$E_{\rm sim}$ values for $\Lambda_h$, $\Sigma_h$, and
$\Sigma^\ast_h$ baryons, and $\Sigma_h^\ast-\Sigma_h$ splittings from ratio fits.
}
\label{tab:baryons}
\end{table}

\begin{table}[thb]
\setlength\tabcolsep{10pt}
\begin{center}
\begin{tabular}{|c|c|c|c|c|c|}
\noalign{\hrule}
\multicolumn{1}{|c}{baryon} &  
\multicolumn{1}{|c}{quark content} &
\multicolumn{2}{|c}{ experimental} &
\multicolumn{1}{|c}{\protect\cite{UKQCDbaryons}} &
\multicolumn{1}{|c|}{Our results} \\
\multicolumn{1}{|c}{} & 
\multicolumn{1}{|c}{} &
\multicolumn{1}{|c}{$c$} &
\multicolumn{1}{|c}{$b$} &
\multicolumn{1}{|c}{$b$} &
\multicolumn{1}{|c|}{$b$} \\
\noalign{\hrule}
\multicolumn{6}{|c|}{$\Lambda$-like ($s_l$ = 0, j= 1/2)} \\
\noalign{\hrule}
$\Lambda_h$      & $(udh)$ & $2.285(1)$                & $5.624(9)$                  & $5.64(^{+5}_{-5})(^{+3}_{-2})$ & $5.679(71)({}^{+14}_{-19})$ \\
$\Xi_h$          & $(lsh)$ & $2.466   $                &                             & $5.76(^{+3}_{-5})(^{+4}_{-3})$ & $5.795(53)({}^{+9}_{-15})(^{+15}_{-0})$ \\
\noalign{\hrule}
\multicolumn{6}{|c|}{$\Sigma$-like ($s_l$ = 1, j= 1/2)} \\
\noalign{\hrule}
$\Sigma_h$       & $(llh)$ & $2.453(1)$             & $5.797(8)$~\protect\cite{lep2} & $5.77(^{+6}_{-6})(^{+4}_{-4})$ & $5.887(49)({}^{+25}_{-37})$ \\
$\Xi_h^\prime$   & $(lsh)$ & $2.574$~\protect\cite{ciulli} &                           & $5.90(^{+6}_{-6})(^{+4}_{-4})$ & $5.968(39)({}^{+20}_{-32})(^{+24}_{-0})$ \\
$\Omega_h$       & $(ssh)$ & $2.704(4)$                &                             & $5.99(^{+5}_{-5})(^{+5}_{-5})$ & $6.048(33)({}^{+16}_{-26})(^{+34}_{-0})$ \\
\noalign{\hrule}
\multicolumn{6}{|c|}{$\Sigma^\ast$-like ($s_l$ = 1, j= 3/2)} \\
\noalign{\hrule}
$\Sigma_h^\ast$  & $(llh)$ & $2.519(2)$             & $5.853(8)$~\protect\cite{lep2} & $5.78(^{+5}_{-6})(^{+4}_{-3})$ & $5.909(47)({}^{+25}_{-39})$ \\
$\Xi_h^\ast$     & $(lsh)$ & $2.645$                   &                             & $5.90(^{+4}_{-6})(^{+4}_{-5})$ & $5.989(39)({}^{+22}_{-34})(^{+25}_{-0})$ \\
$\Omega_h^\ast$  & $(ssh)$ &                           &                             & $6.00(^{+4}_{-5})(^{+5}_{-5})$ & $6.069(34)({}^{+18}_{-30})(^{+35}_{-0})$ \\
\noalign{\hrule}
\end{tabular}
\end{center}
 \par\vfil\penalty-5000\vfilneg
\caption{Summary of masses in GeV for baryons with quark content
shown in column two ($h$ denotes a generic heavy quark ($c$ or $b$),
$l$ stands for a $u$ or $d$ quark). Errors are as explained in the caption to 
Table~\protect\ref{tab:meson_summary}. Finite lattice volume effects, which 
could be large, have not been addressed in this exploratory study. 
Experimental results are given in
columns three and four.  Previous results
(UKQCD~\protect\cite{UKQCDbaryons}) are in column five.  The last
column gives results of our calculation.}
\label{tab:baryexp}
\end{table}

\subsection{\Lb splitting}
\label{ss:LambdaBbar}

\begin{figure}[thbp]
\begin{center}
\epsfysize=3in
\centerline{\epsfbox{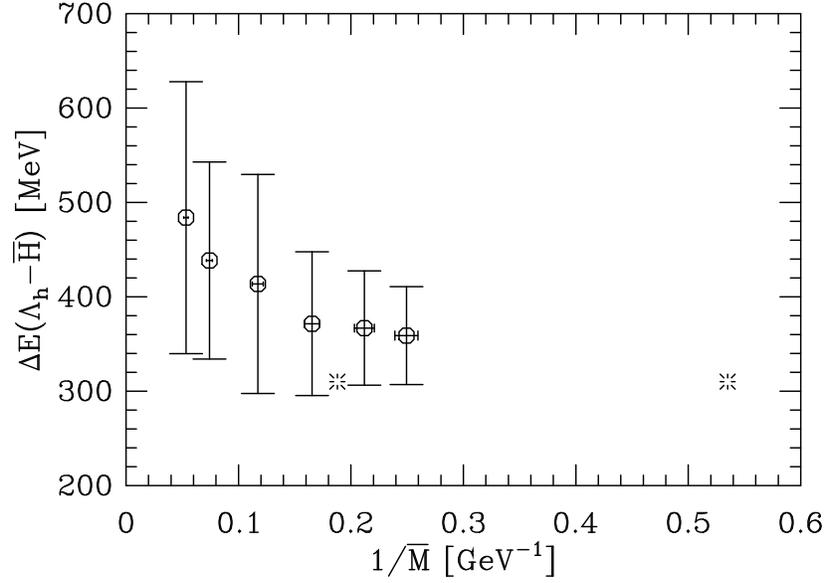 }}
\caption{Spin-averaged $\Lambda_h-H$ splitting as a function of 
$1/\overline{M}$. }
\label{fig:LambdaBbar}
\vspace{-1cm}
\end{center}
\end{figure}

\begin{figure}[thbp]
\begin{center}
\epsfysize=3in
\centerline{\epsfbox{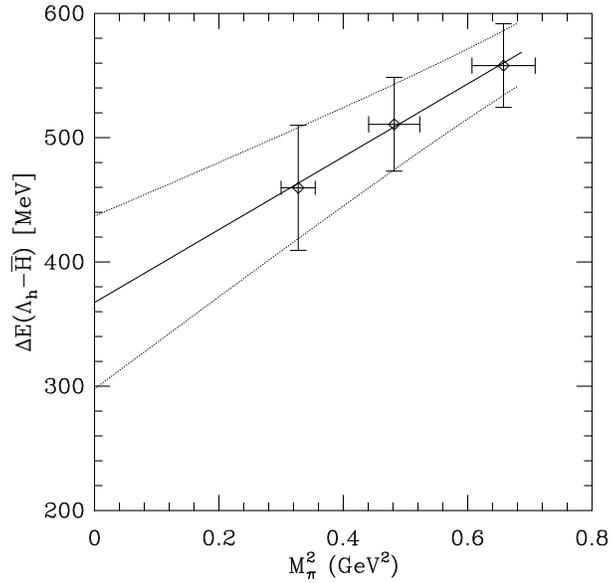 }}
\caption{Spin-averaged $\Lambda_h-H$ splitting as a function of the light quark mass represented 
by the corresponding pseudoscalar meson $M_\pi^2$. }
\label{fig:LambdabBbarvspi2}
\vspace{-1cm}
\end{center}
\end{figure}

We first consider the splitting $ M_{\Lambda_h} - \left(M_H +
3M_{H^\ast}\right)/4 $. In this combination, the heavy quark mass
cancels and there is no contribution from the hyperfine interaction
$E_{\sigma_H \cdot \sigma_l}$.  Since the light quarks are in a ground
state with total spin zero, the mass of the extra light quark in the
baryon gives the dominant contribution. This is borne out by the
experimental values: $311(10)$ and $310(2)$ MeV for the $b$ and $c$
systems respectively, indicating the absence of $O(\Lambda_{QCD}^2 /
M)$ contributions from the difference in kinetic energy to the
splitting (see Fig.~\ref{fig:overview1}). Our lattice data, displayed in 
Table~\ref{tab:LambdaBbar},
show little dependence on the heavy quark mass, Fig.~\ref{fig:LambdaBbar}. 
The variation with the light quark mass is linear as expected,  see
Fig.~\ref{fig:LambdabBbarvspi2}. Our
estimates are $\Lambda_b - \overline{B} = 370(67)({}^{+14}_{-20})$ MeV and
$\Xi_b - \overline{B}_s = 392(50)({}^{+15}_{-0})$ MeV.

There exist a number of previous results for $\Lambda_b - B$, obtained
by extrapolating in the heavy quark mass,
$359({}^{+55}_{-45})({}^{+27}_{-26})$ MeV~\cite{UKQCDbaryons} and
$458(144)(18)$ MeV~\cite{WUPPalexandrou}; in the static limit,
$420({}^{+100}_{-90})({}^{+30}_{-30})$ MeV~\cite{UKQCDmesons}; and with
NRQCD on coarse lattices $363(9)$ MeV~\cite{McKELLARnrqcd} (no systematic errors 
quoted). These 
values are consistent with our result $\Lambda_b - B =
388(68)({}^{+15}_{-23})$ MeV.


\begin{table}[thbp]
\lattrue\interfalse 
 \begingroup
 \begin{center}
 \setupcounts
 \edef\selectcols#1#2#3#4#5#6#7{\selectcols{#1}{#2}{#3}{#4}{#5}{#6}{#7}}
 \edef\putrow#1#2#3#4{\putrow{#1}{#2}{#3}{#4}}
 \begin{tabular}{\putrow{|c|}{r|}%
                 {\selectcols{r|}{r|}{r|}{r|}{r|}{r|}{r|}}%
                 {\selectcols{r|}{r|}{r|}{r|}{r|}{r|}{r|}}}
 \noalign{\hrule}
 \multicolumn{\allcolcount}{|c|}{  }\\[-12pt]
 \multicolumn{\allcolcount}{|c|}{
 \ensuremath{{\Delta E}(\!\Lambda_h-\overline{H})}        }\\
 \noalign{\hrule}
 \multicolumn{1}{|c|}{} &
 \ifchi\multicolumn{1}{|c|}{} &\fi
 \iflat\multicolumn{\colcount}{c|}{lattice units}&\fi
 \ifMeV\multicolumn{\colcount}{|c|}{MeV}\fi\\
 \noalign{\hrule}
 \putrow{$aM^0$}{&\multicolumn{1}{c|}{\ensuremath{\chi^2}}}%
 {\selectcols{&\multicolumn{1}{c|}{\ensuremath{\kappa_{light}          }}}
 {&\multicolumn{1}{c|}{\ensuremath{\kappa_{av}(m_K)        }}}
 {&\multicolumn{1}{c|}{\ensuremath{\kappa_{av}(m_{K^\ast}) }}}
 {&\multicolumn{1}{c|}{\ensuremath{\kappa_{av}(m_{\phi})   }}}
 {&\multicolumn{1}{c|}{\ensuremath{\kappa_{s}(m_K)         }}}
 {&\multicolumn{1}{c|}{\ensuremath{\kappa_{s}(m_{K^\ast})  }}}
 {&\multicolumn{1}{c|}{\ensuremath{\kappa_{s}(m_\phi)      }}}}%
 {\selectcols{&\multicolumn{1}{c|}{\ensuremath{\kappa_{light}          }}}
 {&\multicolumn{1}{c|}{\ensuremath{\kappa_{av}(m_K)        }}}
 {&\multicolumn{1}{c|}{\ensuremath{\kappa_{av}(m_{K^\ast}) }}}
 {&\multicolumn{1}{c|}{\ensuremath{\kappa_{av}(m_{\phi})   }}}
 {&\multicolumn{1}{c|}{\ensuremath{\kappa_{s}(m_K)         }}}
 {&\multicolumn{1}{c|}{\ensuremath{\kappa_{s}(m_{K^\ast})  }}}
 {&\multicolumn{1}{c|}{\ensuremath{\kappa_{s}(m_\phi)      }}}}\\
 \noalign{\hrule}
 \putrow{1.6                                               }{&$0.57E-01$}%
{\selectcols{&$0.187(25)$}{&$0.222(19)$}{&$0.234(17)$}{&$0.234(17)$}{&$0.263(13)$}{&$0.280(15)$}{&$0.281(15)$}}%
{\selectcols{&$ 359( 52)$}{&$ 426( 40)$}{&$ 448( 35)$}{&$ 448( 35)$}{&$ 504( 27)$}{&$ 536( 22)$}{&$ 538( 22)$}}\\
 \putrow{2.0                                               }{&$0.74E-01$}%
{\selectcols{&$0.191(30)$}{&$0.224(22)$}{&$0.235(20)$}{&$0.236(20)$}{&$0.263(14)$}{&$0.280(15)$}{&$0.280(15)$}}%
{\selectcols{&$ 367( 60)$}{&$ 430( 46)$}{&$ 451( 40)$}{&$ 452( 40)$}{&$ 505( 31)$}{&$ 535( 25)$}{&$ 537( 24)$}}\\
 \putrow{2.7                                               }{&$0.46E-01$}%
{\selectcols{&$0.194(38)$}{&$0.226(28)$}{&$0.237(25)$}{&$0.237(25)$}{&$0.264(18)$}{&$0.280(18)$}{&$0.281(17)$}}%
{\selectcols{&$ 372( 76)$}{&$ 433( 58)$}{&$ 454( 51)$}{&$ 455( 51)$}{&$ 506( 38)$}{&$ 536( 31)$}{&$ 538( 30)$}}\\
 \putrow{4.0                                               }{&$0.27E-01$}%
{\selectcols{&$0.215(59)$}{&$0.239(44)$}{&$0.247(39)$}{&$0.247(39)$}{&$0.266(28)$}{&$0.278(22)$}{&$0.278(22)$}}%
{\selectcols{&$ 414(116)$}{&$ 458( 89)$}{&$ 473( 79)$}{&$ 473( 78)$}{&$ 510( 59)$}{&$ 532( 45)$}{&$ 533( 44)$}}\\
 \putrow{7.0                                               }{&$0.27    $}%
{\selectcols{&$0.228(52)$}{&$0.244(37)$}{&$0.250(33)$}{&$0.250(33)$}{&$0.263(28)$}{&$0.271(30)$}{&$0.272(30)$}}%
{\selectcols{&$ 438(104)$}{&$ 469( 77)$}{&$ 479( 69)$}{&$ 480( 69)$}{&$ 505( 60)$}{&$ 520( 60)$}{&$ 521( 60)$}}\\
 \putrow{10.0                                              }{&$0.32    $}%
{\selectcols{&$0.252(72)$}{&$0.258(50)$}{&$0.260(45)$}{&$0.260(44)$}{&$0.264(38)$}{&$0.267(41)$}{&$0.268(41)$}}%
{\selectcols{&$ 484(144)$}{&$ 495(103)$}{&$ 498( 92)$}{&$ 499( 92)$}{&$ 507( 79)$}{&$ 513( 82)$}{&$ 513( 83)$}}\\
 \noalign{\hrule}
 \end{tabular}
 \end{center}
 \endgroup
 \par\vfil\penalty-5000\vfilneg

\caption{Splitting between the $\Lambda_h$ and the spin-averaged $H$. The
experimental value for  $\Lambda_h - \overline{B}_d$ is $310(11)$ MeV.}
\label{tab:LambdaBbar}
\end{table}

\subsection{\sigl splitting}
\label{ss:SigmabarLambda}

\begin{figure}[thbp]
\begin{center}
\epsfysize=3in
\centerline{\epsfbox{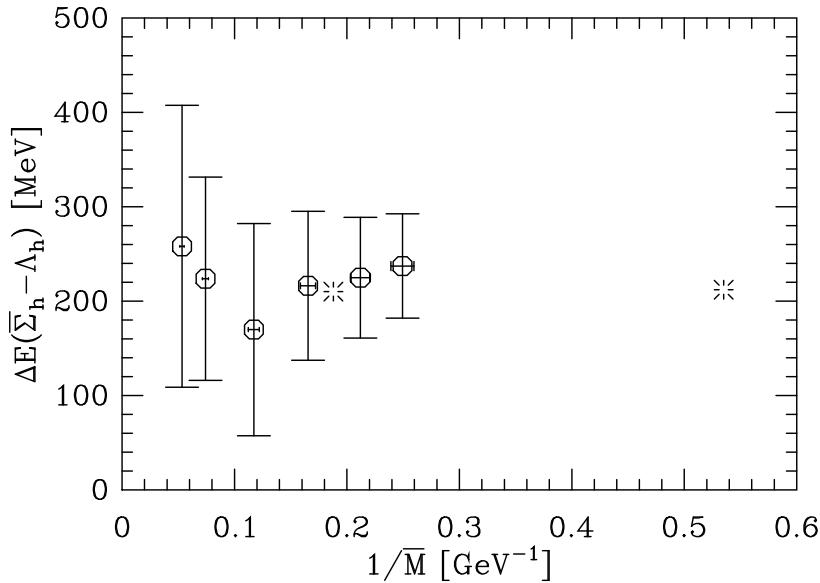 }}
\caption{Spin-averaged $\overline{\Sigma}_h-\Lambda_h$ splitting as a 
function of the inverse spin-averaged meson mass. The bursts denote 
experimental values for $b$ and $c$ heavy quarks.}
\label{fig:SigmabarLambda}
\vspace{-1cm}
\end{center}
\end{figure}

In our picture, the splitting $(2\Sigma_h +
4\Sigma^\ast_h)/6-\Lambda_h$ depends on $E_{\sigma_l \cdot \sigma_l}$, the
hyperfine interaction between the light quarks, the difference of the
binding energies, and of the kinetic energies of the heavy quark in
each baryon. Experimentally, it is found to be independent of
the heavy quark mass: $(2\Sigma_c + 4\Sigma^\ast_c)/6-\Lambda_c = 212$ MeV 
and the preliminary estimate 
$(2\Sigma_b + 4\Sigma^\ast_b)/6-\Lambda_b= 210$ MeV (see also
Fig.~\ref{fig:overview1}). These numbers are roughly $2/3$ of the 
Delta-Nucleon splitting ($293$ MeV). Such a ratio is obtained in a 
simple non-relativistic model where these splittings are dominated by the 
light quark hyperfine interaction. The lattice results
shown in Table~\ref{tab:SigmabarLambda} and Fig.~\ref{fig:SigmabarLambda} 
are also independent of the
heavy quark mass and give $221(71)({}^{+12}_{-16})$ MeV at $M_b$. 

In the charmed sector the experimental value changes
significantly on replacing $d$ with $s$, $i.e.$ $(2\Xi^\prime_c +
4\Xi^\ast_c)/6 - \Xi_c = 154$ MeV.  Our lattice results at the $b$
mass also  show a decrease with $221(71)({}^{+12}_{-16})$ going to 
$186(51)({}^{+13}_{-17})({}^{+0}_{-10})$ MeV, although the difference is
not statistically significant.

The UKQCD collaboration~\cite{UKQCDbaryons} reports $\Sigma_b -
\Lambda_b = 190({}^{+60}_{-75})({}^{+30}_{-30}) $ MeV and
$\Xi^\prime_b - \Xi_b = 157({}^{+52}_{-64})({}^{+11}_{-11})$ MeV from 
extrapolating in the heavy quark mass to the $b$. Our
results for these splittings are $209(71) $ and $177(54)({}^{+0}_{-10})$
MeV respectively.

\begin{table}[thbp]
\lattrue
 \begingroup
 \begin{center}
 \setupcounts
 \edef\selectcols#1#2#3#4#5#6#7{\selectcols{#1}{#2}{#3}{#4}{#5}{#6}{#7}}
 \edef\putrow#1#2#3#4{\putrow{#1}{#2}{#3}{#4}}
 \begin{tabular}{\putrow{|c|}{r|}%
                 {\selectcols{r|}{r|}{r|}{r|}{r|}{r|}{r|}}%
                 {\selectcols{r|}{r|}{r|}{r|}{r|}{r|}{r|}}}
 \noalign{\hrule}
 \multicolumn{\allcolcount}{|c|}{   }\\[-12pt]
 \multicolumn{\allcolcount}{|c|}{
 \ensuremath{{\Delta E}({\overline\Sigma}_h\!-\!\Lambda_h)}   }\\
 \noalign{\hrule}
 \multicolumn{1}{|c|}{} &
 \ifchi\multicolumn{1}{|c|}{} &\fi
 \iflat\multicolumn{\colcount}{c|}{lattice units}&\fi
 \ifMeV\multicolumn{\colcount}{|c|}{MeV}\fi\\
 \noalign{\hrule}
 \putrow{$aM^0$}{&\multicolumn{1}{c|}{\ensuremath{\chi^2}}}%
 {\selectcols{&\multicolumn{1}{c|}{\ensuremath{\kappa_{light}          }}}
 {&\multicolumn{1}{c|}{\ensuremath{\kappa_{av}(m_K)        }}}
 {&\multicolumn{1}{c|}{\ensuremath{\kappa_{av}(m_{K^\ast}) }}}
 {&\multicolumn{1}{c|}{\ensuremath{\kappa_{av}(m_{\phi})   }}}
 {&\multicolumn{1}{c|}{\ensuremath{\kappa_{s}(m_K)         }}}
 {&\multicolumn{1}{c|}{\ensuremath{\kappa_{s}(m_{K^\ast})  }}}
 {&\multicolumn{1}{c|}{\ensuremath{\kappa_{s}(m_\phi)      }}}}%
 {\selectcols{&\multicolumn{1}{c|}{\ensuremath{\kappa_{light}          }}}
 {&\multicolumn{1}{c|}{\ensuremath{\kappa_{av}(m_K)        }}}
 {&\multicolumn{1}{c|}{\ensuremath{\kappa_{av}(m_{K^\ast}) }}}
 {&\multicolumn{1}{c|}{\ensuremath{\kappa_{av}(m_{\phi})   }}}
 {&\multicolumn{1}{c|}{\ensuremath{\kappa_{s}(m_K)         }}}
 {&\multicolumn{1}{c|}{\ensuremath{\kappa_{s}(m_{K^\ast})  }}}
 {&\multicolumn{1}{c|}{\ensuremath{\kappa_{s}(m_\phi)      }}}}\\
 \noalign{\hrule}
 \putrow{1.6                                               }{&$0.69E-01$}%
{\selectcols{&$0.124(29)$}{&$0.102(21)$}{&$0.097(19)$}{&$0.097(19)$}{&$0.080(14\
)$}{&$0.070(14)$}{&$0.070(14)$}}%
{\selectcols{&$ 237( 55)$}{&$ 196( 41)$}{&$ 186( 39)$}{&$ 186( 39)$}{&$ 154( 29\
)$}{&$ 135( 29)$}{&$ 134( 29)$}}\\
 \putrow{2.0                                               }{&$0.66E-01$}%
{\selectcols{&$0.119(34)$}{&$0.099(25)$}{&$0.095(23)$}{&$0.095(23)$}{&$0.080(17\
)$}{&$0.071(15)$}{&$0.071(15)$}}%
{\selectcols{&$ 227( 65)$}{&$ 190( 48)$}{&$ 182( 46)$}{&$ 182( 45)$}{&$ 153( 33\
)$}{&$ 137( 32)$}{&$ 136( 32)$}}\\
 \putrow{2.7                                               }{&$0.46E-01$}%
{\selectcols{&$0.106(38)$}{&$0.092(28)$}{&$0.089(26)$}{&$0.089(25)$}{&$0.078(18\
)$}{&$0.072(16)$}{&$0.072(16)$}}%
{\selectcols{&$ 204( 72)$}{&$ 177( 53)$}{&$ 171( 50)$}{&$ 171( 50)$}{&$ 150( 36\
)$}{&$ 138( 33)$}{&$ 138( 33)$}}\\
 \putrow{4.0                                               }{&$0.65E-01$}%
{\selectcols{&$0.092(59)$}{&$0.088(44)$}{&$0.087(40)$}{&$0.087(40)$}{&$0.084(29\
)$}{&$0.082(23)$}{&$0.082(23)$}}%
{\selectcols{&$ 176(112)$}{&$ 169( 83)$}{&$ 167( 77)$}{&$ 166( 76)$}{&$ 161( 56\
)$}{&$ 157( 45)$}{&$ 157( 45)$}}\\
 \putrow{7.0                                               }{&$0.25    $}%
{\selectcols{&$0.126(56)$}{&$0.118(39)$}{&$0.116(37)$}{&$0.116(36)$}{&$0.110(31\
)$}{&$0.106(32)$}{&$0.106(32)$}}%
{\selectcols{&$ 241(108)$}{&$ 226( 76)$}{&$ 223( 71)$}{&$ 223( 71)$}{&$ 211( 59\
)$}{&$ 204( 62)$}{&$ 204( 62)$}}\\
 \putrow{10.0                                              }{&$0.29    $}%
{\selectcols{&$0.141(79)$}{&$0.135(53)$}{&$0.134(49)$}{&$0.134(49)$}{&$0.129(42\
)$}{&$0.127(45)$}{&$0.127(46)$}}%
{\selectcols{&$ 270(152)$}{&$ 259(102)$}{&$ 257( 95)$}{&$ 256( 95)$}{&$ 248( 81\
)$}{&$ 243( 89)$}{&$ 243( 89)$}}\\
 \noalign{\hrule}
 \end{tabular}
 \end{center}
 \endgroup
 \par\vfil\penalty-5000\vfilneg

\caption{Splitting between the spin-averaged $\Sigma_h$ and $\Lambda_h$ as 
a function of $M^0$. $\kappa_{av}$ corresponds to setting the light quark mass to
$(m_s+m_l)/2$. The preliminary experimental value is $\overline{\Sigma}_b - \Lambda_b = 210$
 MeV~\protect\cite{lep2}.} 
\label{tab:SigmabarLambda}
\end{table}

\subsection{\shyp splitting}
\label{ss:SigmastarSigma}

\begin{table}[thbp]
\lattrue
 \begingroup
 \begin{center}
 \setupcounts
 \edef\selectcols#1#2#3#4#5#6#7{\selectcols{#1}{#2}{#3}{#4}{#5}{#6}{#7}}
 \edef\putrow#1#2#3#4{\putrow{#1}{#2}{#3}{#4}}
 \begin{tabular}{\putrow{|c|}{r|}%
                 {\selectcols{r|}{r|}{r|}{r|}{r|}{r|}{r|}}%
                 {\selectcols{r|}{r|}{r|}{r|}{r|}{r|}{r|}}}
 \noalign{\hrule}
 \multicolumn{\allcolcount}{|c|}{
 \ensuremath{\Delta E(\Sigma^*_h\!-\!\Sigma_h)}   }\\
 \noalign{\hrule}
 \multicolumn{1}{|c|}{} &
 \ifchi\multicolumn{1}{|c|}{} &\fi
 \iflat\multicolumn{\colcount}{c|}{lattice units}&\fi
 \ifMeV\multicolumn{\colcount}{|c|}{MeV}\fi\\
 \noalign{\hrule}
 \putrow{$aM^0$}{&\multicolumn{1}{c|}{\ensuremath{\chi^2}}}%
 {\selectcols{&\multicolumn{1}{c|}{\ensuremath{\kappa_{light}          }}}
 {&\multicolumn{1}{c|}{\ensuremath{\kappa_{av}(m_K)        }}}
 {&\multicolumn{1}{c|}{\ensuremath{\kappa_{av}(m_{K^\ast}) }}}
 {&\multicolumn{1}{c|}{\ensuremath{\kappa_{av}(m_{\phi})   }}}
 {&\multicolumn{1}{c|}{\ensuremath{\kappa_{s}(m_K)         }}}
 {&\multicolumn{1}{c|}{\ensuremath{\kappa_{s}(m_{K^\ast})  }}}
 {&\multicolumn{1}{c|}{\ensuremath{\kappa_{s}(m_\phi)      }}}}%
 {\selectcols{&\multicolumn{1}{c|}{\ensuremath{\kappa_{light}          }}}
 {&\multicolumn{1}{c|}{\ensuremath{\kappa_{av}(m_K)        }}}
 {&\multicolumn{1}{c|}{\ensuremath{\kappa_{av}(m_{K^\ast}) }}}
 {&\multicolumn{1}{c|}{\ensuremath{\kappa_{av}(m_{\phi})   }}}
 {&\multicolumn{1}{c|}{\ensuremath{\kappa_{s}(m_K)         }}}
 {&\multicolumn{1}{c|}{\ensuremath{\kappa_{s}(m_{K^\ast})  }}}
 {&\multicolumn{1}{c|}{\ensuremath{\kappa_{s}(m_\phi)      }}}}\\
 \noalign{\hrule}
 \putrow{1.6                                               }{&$0.73E-02$}%
{\selectcols{&$0.016(03)$}{&$0.015(03)$}{&$0.015(03)$}{&$0.015(03)$}{&$0.014(02\
)$}{&$0.014(02)$}{&$0.014(02)$}}%
{\selectcols{&$  30( 07)$}{&$  28( 05)$}{&$  28( 05)$}{&$  28( 05)$}{&$  27( 04\
)$}{&$  26( 04)$}{&$  26( 04)$}}\\
 \putrow{2.0                                               }{&$0.83E-02$}%
{\selectcols{&$0.012(03)$}{&$0.012(03)$}{&$0.011(03)$}{&$0.011(03)$}{&$0.011(02\
)$}{&$0.011(02)$}{&$0.011(02)$}}%
{\selectcols{&$  23( 07)$}{&$  22( 05)$}{&$  22( 05)$}{&$  22( 05)$}{&$  21( 04\
)$}{&$  21( 04)$}{&$  21( 04)$}}\\
 \putrow{2.7                                               }{&$0.75E-01$}%
{\selectcols{&$0.008(03)$}{&$0.008(02)$}{&$0.008(02)$}{&$0.008(02)$}{&$0.008(02\
)$}{&$0.008(02)$}{&$0.008(02)$}}%
{\selectcols{&$  16( 06)$}{&$  15( 05)$}{&$  15( 04)$}{&$  15( 04)$}{&$  15( 04\
)$}{&$  15( 04)$}{&$  15( 04)$}}\\
 \putrow{4.0                                               }{&$0.10E-01$}%
{\selectcols{&$0.005(03)$}{&$0.005(02)$}{&$0.005(02)$}{&$0.005(02)$}{&$0.005(02\
)$}{&$0.005(02)$}{&$0.005(02)$}}%
{\selectcols{&$   9( 05)$}{&$   9( 04)$}{&$  10( 04)$}{&$  10( 04)$}{&$  10( 03\
)$}{&$  10( 03)$}{&$  10( 03)$}}\\
 \putrow{7.0                                               }{&$0.66E-02$}%
{\selectcols{&$0.003(02)$}{&$0.004(02)$}{&$0.004(02)$}{&$0.004(02)$}{&$0.004(02\
)$}{&$0.004(01)$}{&$0.004(01)$}}%
{\selectcols{&$   6( 05)$}{&$   7( 04)$}{&$   7( 04)$}{&$   7( 04)$}{&$   8( 03\
)$}{&$   8( 03)$}{&$   8( 03)$}}\\
 \putrow{10.0                                              }{&$0.78E-02$}%
{\selectcols{&$0.002(02)$}{&$0.002(02)$}{&$0.002(02)$}{&$0.002(02)$}{&$0.002(02\
)$}{&$0.003(01)$}{&$0.003(01)$}}%
{\selectcols{&$   3( 05)$}{&$   4( 04)$}{&$   4( 04)$}{&$   4( 04)$}{&$   5( 03\
)$}{&$   5( 03)$}{&$   5( 03)$}}\\
 \noalign{\hrule}
 \end{tabular}
 \end{center}
 \endgroup
 \par\vfil\penalty-5000\vfilneg

\caption{$\Sigma^\ast_h -\Sigma_h$ splitting. The preliminary experimental
value for $\Sigma^\ast_b -\Sigma_b$ is $56(8)$ MeV~\protect\cite{lep2}.}
\label{tab:SigmastarSigma}
\end{table}

\begin{figure}[thbp]
\begin{center}
\epsfysize=3in
\centerline{\epsfbox{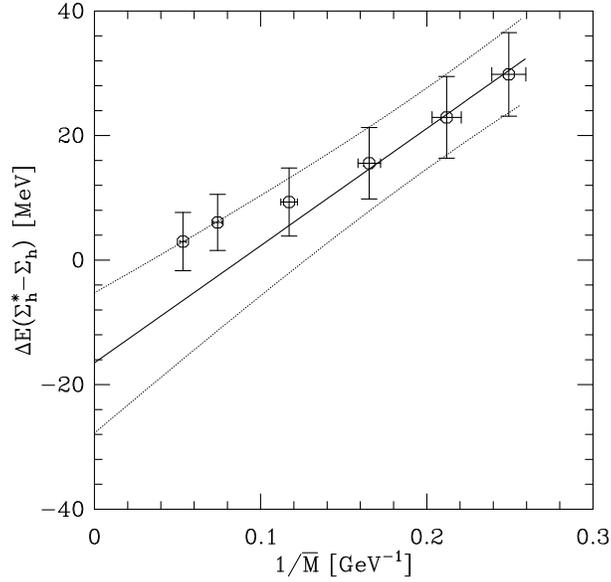 }}
\caption{ $\Sigma^\ast_h-\Sigma_h$ splitting as a 
function of $1/\overline{M}$. }
\label{fig:SigmastarSigma}
\vspace{-1cm}
\end{center}
\end{figure}

The $\Sigma_h^\ast-\Sigma_h$ splitting should depend only on the
heavy-light hyperfine interaction $E_{\sigma_h \cdot \sigma_l}$. It is
therefore expected to be proportional to $1/M_h$. Our lattice
results, shown in Table~\ref{tab:SigmastarSigma}, resolve these splittings for 
the first time. A linear fit
to the three lightest $\overline{M}$ values that bracket $M_b^0$ gives 
$-17(11)({}^{+0}_{-1})$ MeV for the intercept and  
$0.188(44)({}^{+17}_{-22}) {\rm GeV}^2$ for the slope.  However, as
apparent from Fig.~\ref{fig:SigmastarSigma}, if the fit is constrained
to have zero intercept, then it would have a much smaller slope. Based
on the assumption that the wavefunction at the origin is similar, one
expects the slope for the baryon splitting to be 0.75 that for mesons
\cite{rosner96}, which was found to be $0.138(38)({}^{+11}_{-17}) {\rm
GeV}^2$ in Sec.~\ref{ss:BstarB}. This expectation does not hold in
the charm sector where $\Sigma^\ast_c-\Sigma_c \approx 66$ MeV whereas
$D^\ast - D \approx 140$ MeV.


The preliminary experimental value is $\Sigma_b^* - \Sigma_b = 56(8)$
MeV~\cite{lep2}. It is however likely that at least one of the states
has been misidentified~\cite{falk97}, and this number is too
large. Scaling the experimental value $\Sigma^\ast_c-\Sigma_c = 66 $
MeV by $M_c/M_b$ suggests $\sim 20$ MeV for this
splitting~\cite{falk97,jenkins97}.  We find
$\Sigma_b^\ast-\Sigma_b = 19(7)({}^{+2}_{-3})$ MeV; however this could be an
underestimate based on the general discussion of hyperfine interactions in 
Sec.~\ref{ss:BstarB}. 

The raw lattice data does not show a dependence on the light quark
mass.  Experimentally, there exists data for strange baryons only in
the $c$ sector. The preliminary estimate $\Xi^\ast_c - \Xi^\prime_c
\approx 77$ MeV is $\approx 11$ MeV larger than the $\Sigma_c^\ast -
\Sigma_c$ splitting.  At the $b$, heavy quark scaling suggests that this difference should be reduced 
by the factor $M_c / M_b \approx 0.3$, making it much smaller than our
resolution. We find $\Xi_b^\ast - \Xi_b^{\prime} =
19(5)({}^{+2}_{-3})$ and $\Omega_b^\ast - \Omega_b =
18(4)({}^{+2}_{-3})$ MeV.

\section{Heavy-heavy-light baryons}
\label{sec:HHLbaryons}

It is theoretically interesting to study heavy-heavy-light baryons
even though it is exceedingly hard to produce two overlapping $b$
quarks in experiments.  The two heavy quarks are expected to bind in a
color anti-triplet state whose size is much smaller than
$\Lambda_{QCD}$. It thus interacts with
the light degrees of freedom to yield a level structure similar to
that of heavy-light mesons~\cite{savage90,falk94}.

In the $S$-wave baryons, the total angular momentum of the two heavy quarks is
$J=0$ or $1$. For identical quarks only $J=1$ is possible. There are
two different ways to couple the light quark spin to this
configuration. The $J=3/2$ states are denoted as $\Xi^{\ast0}_{bb}$, 
$\Xi^{\ast-}_{bb}$, and $\Omega^{\ast-}_{bb}$, and the $J=1/2$ states as
$ \Xi^{0}_{bb}$, $\Xi^{-}_{bb}$, and $\Omega^{-}_{bb}$ (the quark
content is $bbu$, $bbd$, and $bbs$ respectively). These are split by a
heavy-light hyperfine interaction. Two heavy quarks with different flavor can 
also be in a $J=0$ state, and the corresponding baryons are denoted by us
as $ \Xi^{\prime0}_{bb^\prime}$, $\Xi^{\prime-}_{bb^\prime}$, and
$\Omega^{\prime-}_{bb^\prime}$. The splitting between the spin
averaged $\Xi_{bb}$ and the $\Xi^{\prime}_{bb^\prime}$ (and the
corresponding splitting between the $\Omega$'s) is due to the
heavy-heavy spin interaction. This is expected to be very small,
and to vanish in the infinite mass limit.

\begin{table}[ht]
 \begin{center}
 \begin{tabular}{|c|c|c|c|c|}
 \noalign{\hrule}
 $aM^0$ & $\kappa$&\ensuremath{aE_{\rm sim}(\Xi_{hh})}&
 \ensuremath{a\Delta E(\Xi^\ast_{hh}-\Xi_{hh})} & 
\ensuremath{a \Delta E(\Xi^{\prime}_{hh^\prime}-\Xi_{hh})} \\
\hline
  1.6& 0.13690&0.767( 08)&0.015(02)&0.005(01)\\
  2.0&        &0.788( 10)&0.012(02)&0.004(01)\\
  2.7&        &0.803( 10)&0.009(02)&0.003(01)\\
  4.0&        &0.803( 14)&0.007(02)&0.002(01)\\
  7.0&        &0.767( 34)&0.003(02)&0.001(01)\\
 10.0&        &0.735( 86)&0.001(02)&0.000(01)\\
\hline
  1.6& 0.13750&0.754( 10)&0.015(02)&0.006(02)\\
  2.0&        &0.777( 10)&0.012(02)&0.004(02)\\
  2.7&        &0.788( 12)&0.009(02)&0.003(02)\\
  4.0&        &0.792( 15)&0.007(02)&0.002(01)\\
  7.0&        &0.746( 38)&0.003(02)&0.001(01)\\
 10.0&        &0.717(107)&0.001(02)&0.000(01)\\
\hline
  1.6& 0.13808&0.747( 12)&0.015(02)&0.006(02)\\
  2.0&        &0.768( 14)&0.012(02)&0.004(02)\\
  2.7&        &0.779( 14)&0.009(02)&0.003(02)\\
  4.0&        &0.774( 19)&0.006(02)&0.002(01)\\
  7.0&        &0.733( 47)&0.002(02)&0.001(01)\\
 10.0&        &0.718(144)&$-.001(03)$&0.000(01)\\
 \noalign{\hrule}
 \end{tabular}
 \end{center}
 \par\vfil\penalty-5000\vfilneg
\caption{$E_{\rm sim}$ and splittings for heavy-heavy-light baryons.}
\label{tab:HHLbaryons}
\end{table}

\begin{figure}[thbp]
\begin{center}
\epsfysize=3in
\centerline{\epsfbox{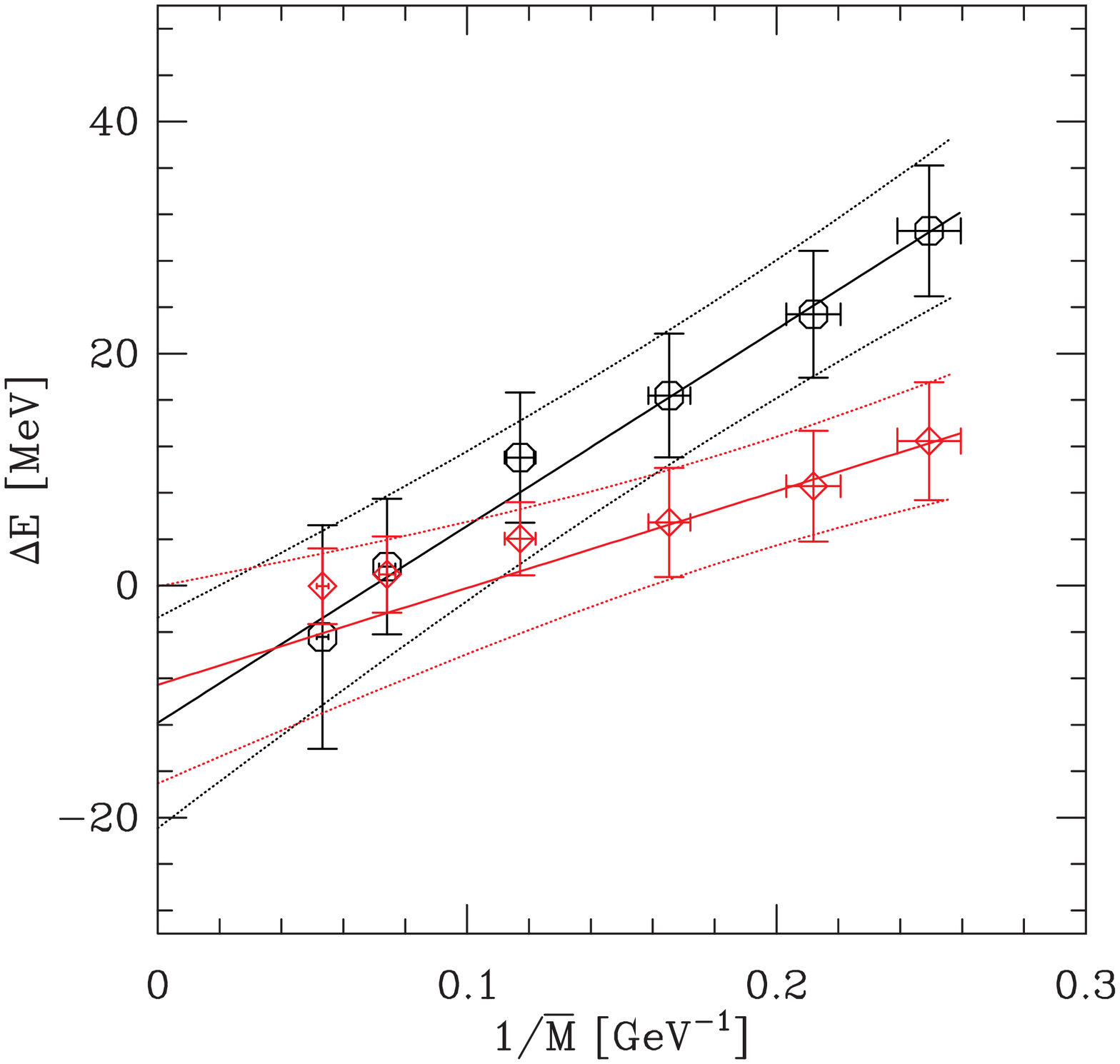 }}
\caption{ $\Xi^\ast_{hh}-\Xi_{hh}$ (circles) and
$\Xi^{\prime}_{hh^\prime}-\Xi_{hh}$ (diamonds) splittings as a 
function of the inverse spin-averaged meson mass along with linear fits. }
\label{fig:SigmastarbbSigmabb}
\vspace{-1cm}
\end{center}
\end{figure}

Our raw data are given in Table~\ref{tab:HHLbaryons}, and the results
for $\Xi^\ast_{hh}-\Xi_{hh}$, extrapolated to $m_l$ and $m_s$, are
listed in Table~\ref{tab:SigmastarbbSigmabb}. The data show a strong
dependence on the heavy quark mass and almost none on the light quark
mass. The slope with respect to $1/\overline{M}$ is
$0.170(42)({}^{+14}_{-21}) {\rm GeV}^2$ as shown in
Fig.~\ref{fig:SigmastarbbSigmabb}, and the intercept is
$-12(9)({}^{+0}_{-1})$ MeV.  These results are consistent with those
for $\Sigma_h^* - \Sigma_h$. Both are hyperfine splittings between
$S=1$ diquark and $S=1/2$ quark sub-systems; the difference is whether the $S=1$
sub-system is heavy-heavy or light-light.  
In principle the strength
of the spin-spin interaction could be different, however, the data suggest 
that they are similar.  In fact this similarity persists even for $h= s$ 
where $\Xi^\ast-\Xi = 210$ MeV and $\Sigma^* - \Sigma = 196$ MeV. 
%

If we assume that the spin interaction between the heavy quarks is negligible, 
then we expect $(\Xi^\ast_{hh}-\Xi_{hh}) = 1.5 (\Xi^\prime_{hh'}-\Xi_{hh})$. 
The data shown in Table~\ref{tab:HHLbaryons} indicates a ratio of three instead. 
Our final estimates are
\begin{eqnarray}
\Xi_{bb} = 10314(46)({}^{-10}_{+11})\MeV,&\qquad
&\Omega_{bb} = 10365(40)({}^{-11}_{+12})(^{+16}_{-0})\MeV,
   \nonumber\\
\Xi^*_{bb} = 10333(55)({}^{-7}_{+6})\MeV,&\qquad
&
\Omega^*_{bb} = 10383(39)({}^{-8}_{+8})(^{+12}_{-0})\MeV,
   \nonumber\\
\Xi^*_{bb}-\Xi_{bb} = 20(6)({}^{+2}_{-3})\MeV,&\qquad
&\Omega^*_{bb}-\Omega_{bb} = 20(4)({}^{+2}_{-3})\MeV. 
   \nonumber
\end{eqnarray}

\begin{table}[thbp]
\lattrue\interfalse 
 \begingroup
 \begin{center}
 \setupcounts
 \edef\selectcols#1#2#3#4#5#6#7{\selectcols{#1}{#2}{#3}{#4}{#5}{#6}{#7}}
 \edef\putrow#1#2#3#4{\putrow{#1}{#2}{#3}{#4}}
 \begin{tabular}{\putrow{|c|}{r|}%
                 {\selectcols{r|}{r|}{r|}{r|}{r|}{r|}{r|}}%
                 {\selectcols{r|}{r|}{r|}{r|}{r|}{r|}{r|}}}
 \noalign{\hrule}
 \multicolumn{\allcolcount}{|c|}{
 \ensuremath{\Xi^\ast_{hh}-\Xi_{hh}}         }\\
 \noalign{\hrule}
 \multicolumn{1}{|c|}{} &
 \ifchi\multicolumn{1}{|c|}{} &\fi
 \iflat\multicolumn{\colcount}{c|}{lattice units}&\fi
 \ifMeV\multicolumn{\colcount}{|c|}{MeV}\fi\\
 \noalign{\hrule}
 \putrow{$aM^0$}{&\multicolumn{1}{c|}{\ensuremath{\chi^2}}}%
 {\selectcols{&\multicolumn{1}{c|}{\ensuremath{\kappa_{light}          }}}
 {&\multicolumn{1}{c|}{\ensuremath{\kappa_{av}(m_K)        }}}
 {&\multicolumn{1}{c|}{\ensuremath{\kappa_{av}(m_{K^\ast}) }}}
 {&\multicolumn{1}{c|}{\ensuremath{\kappa_{av}(m_{\phi})   }}}
 {&\multicolumn{1}{c|}{\ensuremath{\kappa_{s}(m_K)         }}}
 {&\multicolumn{1}{c|}{\ensuremath{\kappa_{s}(m_{K^\ast})  }}}
 {&\multicolumn{1}{c|}{\ensuremath{\kappa_{s}(m_\phi)      }}}}%
 {\selectcols{&\multicolumn{1}{c|}{\ensuremath{\kappa_{light}          }}}
 {&\multicolumn{1}{c|}{\ensuremath{\kappa_{av}(m_K)        }}}
 {&\multicolumn{1}{c|}{\ensuremath{\kappa_{av}(m_{K^\ast}) }}}
 {&\multicolumn{1}{c|}{\ensuremath{\kappa_{av}(m_{\phi})   }}}
 {&\multicolumn{1}{c|}{\ensuremath{\kappa_{s}(m_K)         }}}
 {&\multicolumn{1}{c|}{\ensuremath{\kappa_{s}(m_{K^\ast})  }}}
 {&\multicolumn{1}{c|}{\ensuremath{\kappa_{s}(m_\phi)      }}}}\\
 \noalign{\hrule}
 \putrow{1.6                                               }{&$0.38E-02$}%
{\selectcols{&$0.016( 03)$}{&$0.016( 02)$}{&$0.015( 02)$}{&$0.015( 02)$}{&$0.015( 02)$}{&$0.015( 02)$}{&$0.015( 02)$}}%
{\selectcols{&$  31( 06)$}{&$  30( 05)$}{&$  30( 05)$}{&$  30( 05)$}{&$  29( 04)$}{&$  29( 04)$}{&$  29( 04)$}}\\
 \putrow{2.0                                               }{&$0.36E-02$}%
{\selectcols{&$0.012( 03)$}{&$0.012( 02)$}{&$0.012( 02)$}{&$0.012( 02)$}{&$0.012( 02)$}{&$0.012( 02)$}{&$0.012( 02)$}}%
{\selectcols{&$  23( 05)$}{&$  23( 05)$}{&$  23( 04)$}{&$  23( 04)$}{&$  23( 04)$}{&$  23( 04)$}{&$  23( 04)$}}\\
 \putrow{2.7                                               }{&$0.36E-02$}%
{\selectcols{&$0.009( 03)$}{&$0.009( 02)$}{&$0.009( 02)$}{&$0.009( 02)$}{&$0.009( 02)$}{&$0.009( 02)$}{&$0.009( 02)$}}%
{\selectcols{&$  16( 05)$}{&$  17( 04)$}{&$  17( 04)$}{&$  17( 04)$}{&$  17( 04)$}{&$  17( 03)$}{&$  17( 03)$}}\\
 \putrow{4.0                                               }{&$0.45E-02$}%
{\selectcols{&$0.006( 03)$}{&$0.006( 03)$}{&$0.006( 02)$}{&$0.006( 02)$}{&$0.007( 02)$}{&$0.007( 02)$}{&$0.007( 02)$}}%
{\selectcols{&$  11( 06)$}{&$  12( 05)$}{&$  12( 05)$}{&$  12( 05)$}{&$  13( 04)$}{&$  13( 04)$}{&$  13( 04)$}}\\
 \putrow{7.0                                               }{&$0.74E-02$}%
{\selectcols{&$0.001( 03)$}{&$0.002( 03)$}{&$0.002( 02)$}{&$0.002( 02)$}{&$0.002( 02)$}{&$0.003( 02)$}{&$0.003( 02)$}}%
{\selectcols{&$   2( 06)$}{&$   3( 05)$}{&$   3( 05)$}{&$   3( 05)$}{&$   5( 04)$}{&$   5( 04)$}{&$   5( 04)$}}\\
 \putrow{10.0                                              }{&$0.49    $}%
{\selectcols{&$-.002( 05)$}{&$-.001( 03)$}{&$-.001( 03)$}{&$-.001( 03)$}{&$0.000( 02)$}{&$0.001( 02)$}{&$0.001( 02)$}}%
{\selectcols{&$  -4( 10)$}{&$  -2( 06)$}{&$  -2( 05)$}{&$  -2( 05)$}{&$   0( 04)$}{&$   1( 04)$}{&$   1( 04)$}}\\
 \noalign{\hrule}
 \end{tabular}
 \end{center}
 \endgroup
 \par\vfil\penalty-5000\vfilneg

\caption{$\Xi^\ast_{hh} -\Xi_{hh}$ splitting.}
\label{tab:SigmastarbbSigmabb}
\end{table}

\section{Determination of HQET parameters}
\label{sec:HQET}

We now present a determination of the HQET parameters
$\overline{\Lambda}$, $\lambda_1$, and $\lambda_2$. 
$\overline{\Lambda}$ denotes the binding energy of the meson in the limit $M^0 = \infty$. 
In the static theory the $O(1/M)$ corrections to this are given by 
the expectation value of the heavy quark $p^2$:
\begin{equation}
-\lambda_1 = \frac{1}{2M_B} \langle B|\bar{b}(i\vec{D})^2b|B\rangle,
\end{equation}
and the expectation value of the chromomagnetic operator:
\begin{equation}
\lambda_2 =- 
\frac{1}{2M_B} \langle B|\bar{b}\vec{\sigma}\cdot\vec{B} b|B\rangle.
\end{equation}
Thus, to $O(1/M)$, the relation between the heavy quark pole mass $m_{pole}$ and the heavy-light
meson mass is given by:
\begin{equation}
M_B = m_{pole} + \overline{\Lambda} + \frac{1}{2m_{pole}}\left(-\lambda_1 + \lambda_2\right) 
    \equiv  m_{pole} + E_{\rm bind} \,.
\end{equation}
In NRQCD one measures $E_{\rm sim}$, from which $E_{\rm bind}$ is obtained as 
\begin{equation}
E_{\rm bind} =  E_{\rm sim} - E_0 \,.
\end{equation}
Using the estimates for $E_0$ given in Table~\ref{tab:pert}, $E_{\rm bind}$ 
for the spin-averaged $H$
meson is given in Table~\ref{tab:LambdabarB}, and for $\Lambda_h$ in
Table~\ref{tab:LambdabarL}. 

\begin{figure}[thbp]
\begin{center}
\epsfysize=3in
\centerline{\epsfbox{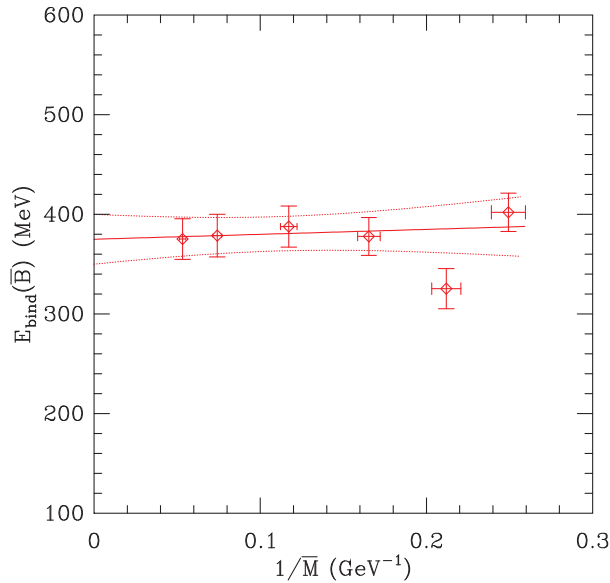 }}
\caption{ $E_{\rm bind}$ versus $1/\overline M$.}
\label{fig:Lambdabar}
\vspace{-1cm}
\end{center}
\end{figure}

We prefer to analyze the dependence of $E_{\rm bind}$ on the heavy
quark mass in terms of $\overline{M}$.  The reason for this choice is
that $m_{pole}$ is not a physical (measurable) quantity and suffers
from a renormalon ambiguity. Also, to $O(1/M)$ the change from
$m_{pole}$ to $\overline{M}$ is benign, $i.e.$, the slope still gives
the same $\lambda_1$ and $\lambda_2$ as extracted in conventional HQET
analyses. The data for the binding energy for $\overline{H}$, and fits
versus $\overline{M}$ are shown in Fig.~\ref{fig:Lambdabar}.  The
behavior of $\Lambda_h$ is similar. The results for
$\overline{\Lambda}$ and $\lambda_1$ obtained from these fits are also
given in Tables~\ref{tab:LambdabarB} and~\ref{tab:LambdabarL}.  Note
that the slope for spin-averaged cases gives $\lambda_1$ since there is
no contribution from the chromomagnetic operator.

Our definition of the parameters $\overline{\Lambda}$ and $\lambda_1$
is perturbative and they inherit a renormalon ambiguity from $E_0$ which could
be as large as $O(\Lambda_{QCD})$ in $\overline \Lambda$. With 
this definition, estimates of the HQET parameters are 
\begin{eqnarray}
  \overline{\Lambda}(B) = 375(25)(50)({}^{+16}_{-22}) {\rm MeV}, &&
    \qquad -\lambda_1(B) = 0.1(3)(1)({}^{+1}_{-1})
      {\rm GeV}^2, \nonumber \\
  \overline{\Lambda}(\Lambda_b) = 895(218)(50)({}^{+37}_{-56}) {\rm MeV}, &&
    \qquad -\lambda_1(\Lambda_b) = -1.7(34)(1)({}^{+2}_{-2})
      {\rm GeV}^2 \,.  
\end{eqnarray}
We have quoted, as the second error, a systematic uncertainty due to
the unknown $O(\alpha_s^2)$ error in the perturbative expansion of
$E_0$, which we take to be $1 \times \alpha_s^2$.  The third
error is due to the scale uncertainty. We emphasize that, due to the
renormalon ambiguity, these estimates are only meant to be indicative
and cannot be compared directly with other calculations.

\begin{table}[ht]
\interfalse 
 \begingroup
 \MeVfalse
 \begin{center}
 \setupcounts
 \edef\selectcols#1#2#3#4#5#6#7{\selectcols{#1}{#2}{#3}{#4}{#5}{#6}{#7}}
 \edef\putrow#1#2#3#4{\putrow{#1}{#2}{#3}{#4}}
 \begin{tabular}{\putrow{|c|}{r|}%
                 {\selectcols{r|}{r|}{r|}{r|}{r|}{r|}{r|}}%
                 {\selectcols{r|}{r|}{r|}{r|}{r|}{r|}{r|}}}
 \noalign{\hrule}
 \multicolumn{\allcolcount}{|c|}{
 \ensuremath{E_{\rm bind}(\overline{H})                }}\\
 \noalign{\hrule}
 \putrow{$aM^0$}{&\multicolumn{1}{c|}{\ensuremath{\chi^2}}}%
 {\selectcols{&\multicolumn{1}{c|}{\ensuremath{\kappa_{light}          }}}
 {&\multicolumn{1}{c|}{\ensuremath{\kappa_{av}(m_K)        }}}
 {&\multicolumn{1}{c|}{\ensuremath{\kappa_{av}(m_{K^\ast}) }}}
 {&\multicolumn{1}{c|}{\ensuremath{\kappa_{av}(m_{\phi})   }}}
 {&\multicolumn{1}{c|}{\ensuremath{\kappa_{s}(m_K)         }}}
 {&\multicolumn{1}{c|}{\ensuremath{\kappa_{s}(m_{K^\ast})  }}}
 {&\multicolumn{1}{c|}{\ensuremath{\kappa_{s}(m_\phi)      }}}}%
 {\selectcols{&\multicolumn{1}{c|}{\ensuremath{\kappa_{light}          }}}
 {&\multicolumn{1}{c|}{\ensuremath{\kappa_{av}(m_K)        }}}
 {&\multicolumn{1}{c|}{\ensuremath{\kappa_{av}(m_{K^\ast}) }}}
 {&\multicolumn{1}{c|}{\ensuremath{\kappa_{av}(m_{\phi})   }}}
 {&\multicolumn{1}{c|}{\ensuremath{\kappa_{s}(m_K)         }}}
 {&\multicolumn{1}{c|}{\ensuremath{\kappa_{s}(m_{K^\ast})  }}}
 {&\multicolumn{1}{c|}{\ensuremath{\kappa_{s}(m_\phi)      }}}}\\
 \noalign{\hrule}
 \putrow{1.6                                               }{&$0.24    $}%
{\selectcols{&$ 402( 19)$}{&$ 445( 16)$}{&$ 459( 13)$}{&$ 460( 13)$}{&$ 496( 14)$}{&$ 516( 09)$}{&$ 517( 09)$}}%
{\selectcols{&$ 462( 21)$}{&$ 505( 18)$}{&$ 519( 15)$}{&$ 520( 15)$}{&$ 556( 16)$}{&$ 576( 10)$}{&$ 577( 10)$}}\\
 \putrow{2.0                                               }{&$0.70E-01$}%
{\selectcols{&$ 325( 20)$}{&$ 368( 15)$}{&$ 382( 12)$}{&$ 383( 12)$}{&$ 418( 11)$}{&$ 439( 07)$}{&$ 440( 07)$}}%
{\selectcols{&$ 410( 22)$}{&$ 453( 18)$}{&$ 467( 15)$}{&$ 468( 14)$}{&$ 503( 14)$}{&$ 524( 08)$}{&$ 525( 08)$}}\\
 \putrow{2.7                                               }{&$0.15    $}%
{\selectcols{&$ 378( 19)$}{&$ 418( 16)$}{&$ 432( 13)$}{&$ 432( 13)$}{&$ 466( 13)$}{&$ 485( 09)$}{&$ 486( 09)$}}%
{\selectcols{&$ 364( 19)$}{&$ 404( 15)$}{&$ 417( 13)$}{&$ 418( 13)$}{&$ 452( 13)$}{&$ 471( 08)$}{&$ 472( 08)$}}\\
 \putrow{4.0                                               }{&$0.93E-01$}%
{\selectcols{&$ 388( 21)$}{&$ 428( 17)$}{&$ 442( 14)$}{&$ 442( 14)$}{&$ 476( 14)$}{&$ 496( 09)$}{&$ 497( 09)$}}%
{\selectcols{&$ 315( 19)$}{&$ 356( 15)$}{&$ 369( 12)$}{&$ 370( 12)$}{&$ 404( 12)$}{&$ 423( 08)$}{&$ 424( 08)$}}\\
 \putrow{7.0                                               }{&$0.12E-01$}%
{\selectcols{&$ 379( 21)$}{&$ 419( 17)$}{&$ 433( 14)$}{&$ 433( 14)$}{&$ 467( 13)$}{&$ 487( 10)$}{&$ 487( 10)$}}%
{\selectcols{&$ 274( 20)$}{&$ 315( 14)$}{&$ 328( 12)$}{&$ 328( 12)$}{&$ 363( 10)$}{&$ 382( 09)$}{&$ 383( 09)$}}\\
 \putrow{10.0                                              }{&$0.15E-01$}%
{\selectcols{&$ 375( 20)$}{&$ 414( 16)$}{&$ 427( 14)$}{&$ 427( 14)$}{&$ 460( 13)$}{&$ 478( 09)$}{&$ 479( 09)$}}%
{\selectcols{&$ 262( 18)$}{&$ 301( 13)$}{&$ 313( 12)$}{&$ 314( 12)$}{&$ 346( 10)$}{&$ 365( 08)$}{&$ 366( 08)$}}\\
 \noalign{\hrule}
 \putrow{}{&}{\selectcols{&}{&}{&}{&}{&}{&}{&}}{\selectcols{&}{&}{&}{&}{&}{&}{&}}\\[-10pt]
 \putrow{\ensuremath{\overline\Lambda}                     }{&          }%
{\selectcols{&$ 375( 25)$}{&$ 414( 19)$}{&$ 426( 16)$}{&$ 426( 16)$}{&$ 458( 14)$}{&$ 477( 11)$}{&$ 477( 11)$}}%
{\selectcols{&$ 203( 18)$}{&$ 241( 13)$}{&$ 254( 12)$}{&$ 254( 12)$}{&$ 286(  9)$}{&$ 304(  9)$}{&$ 305(  9)$}}\\
 \putrow{\ensuremath{-\lambda_1}                      }{&          }%
{\selectcols{& 0.10(33)}{& 0.13(23)}{& 0.14(20)}{& 0.14(20)}{& 0.18(14)}{& 0.20(13)}{& 0.20(13)}}%
{\selectcols{& 2.02(26)}{& 2.05(22)}{& 2.06(21)}{& 2.06(21)}{& 2.09(19)}{& 2.11(19)}{& 2.11(19)}}\\
 \noalign{\hrule}
 \end{tabular}
 \end{center}
 \endgroup
 \par\vfil\penalty-5000\vfilneg

\caption{Binding energies in MeV for the spin-averaged $H$ meson.}
\label{tab:LambdabarB}
\end{table}

\begin{table}[ht]
\strangefalse 
 \begingroup
 \MeVfalse
 \begin{center}
 \setupcounts
 \edef\selectcols#1#2#3#4#5#6#7{\selectcols{#1}{#2}{#3}{#4}{#5}{#6}{#7}}
 \edef\putrow#1#2#3#4{\putrow{#1}{#2}{#3}{#4}}
 \begin{tabular}{\putrow{|c|}{r|}%
                 {\selectcols{r|}{r|}{r|}{r|}{r|}{r|}{r|}}%
                 {\selectcols{r|}{r|}{r|}{r|}{r|}{r|}{r|}}}
 \noalign{\hrule}
 \multicolumn{\allcolcount}{|c|}{
 \ensuremath{E_{\rm bind}(\Lambda_h)}        }\\
 \noalign{\hrule}
 \putrow{$m_Q^0$}{&\multicolumn{1}{c|}{\ensuremath{\chi^2}}}%
 {\selectcols{&\multicolumn{1}{c|}{\ensuremath{\kappa_{light}          }}}
 {&\multicolumn{1}{c|}{\ensuremath{\kappa_{av}(m_K)        }}}
 {&\multicolumn{1}{c|}{\ensuremath{\kappa_{av}(m_{K^\ast}) }}}
 {&\multicolumn{1}{c|}{\ensuremath{\kappa_{av}(m_{\phi})   }}}
 {&\multicolumn{1}{c|}{\ensuremath{\kappa_{s}(m_K)         }}}
 {&\multicolumn{1}{c|}{\ensuremath{\kappa_{s}(m_{K^\ast})  }}}
 {&\multicolumn{1}{c|}{\ensuremath{\kappa_{s}(m_\phi)      }}}}%
 {\selectcols{&\multicolumn{1}{c|}{\ensuremath{\kappa_{light}          }}}
 {&\multicolumn{1}{c|}{\ensuremath{\kappa_{av}(m_K)        }}}
 {&\multicolumn{1}{c|}{\ensuremath{\kappa_{av}(m_{K^\ast}) }}}
 {&\multicolumn{1}{c|}{\ensuremath{\kappa_{av}(m_{\phi})   }}}
 {&\multicolumn{1}{c|}{\ensuremath{\kappa_{s}(m_K)         }}}
 {&\multicolumn{1}{c|}{\ensuremath{\kappa_{s}(m_{K^\ast})  }}}
 {&\multicolumn{1}{c|}{\ensuremath{\kappa_{s}(m_\phi)      }}}}\\
 \noalign{\hrule}
 \putrow{1.6                                               }{&$0.26    $}%
{\selectcols{&$ 761( 60)$}{&$ 870( 47)$}{&$ 907( 40)$}{&$ 908( 40)$}{&$1000( 35)$}{&$1053( 25)$}{&$1055( 25)$}}%
{\selectcols{&$ 821( 61)$}{&$ 930( 49)$}{&$ 967( 41)$}{&$ 968( 41)$}{&$1060( 37)$}{&$1113( 26)$}{&$1115( 25)$}}\\
 \putrow{2.0                                               }{&$0.26    $}%
{\selectcols{&$ 692( 66)$}{&$ 798( 52)$}{&$ 833( 45)$}{&$ 835( 44)$}{&$ 923( 37)$}{&$ 974( 27)$}{&$ 977( 27)$}}%
{\selectcols{&$ 777( 68)$}{&$ 883( 54)$}{&$ 918( 46)$}{&$ 919( 46)$}{&$1008( 39)$}{&$1059( 28)$}{&$1062( 28)$}}\\
 \putrow{2.7                                               }{&$0.12    $}%
{\selectcols{&$ 749( 80)$}{&$ 852( 62)$}{&$ 885( 54)$}{&$ 887( 54)$}{&$ 972( 44)$}{&$1022( 33)$}{&$1024( 33)$}}%
{\selectcols{&$ 735( 80)$}{&$ 837( 62)$}{&$ 871( 54)$}{&$ 872( 54)$}{&$ 958( 44)$}{&$1008( 33)$}{&$1010( 33)$}}\\
 \putrow{4.0                                               }{&$0.12E-01$}%
{\selectcols{&$ 801(120)$}{&$ 886( 94)$}{&$ 915( 82)$}{&$ 916( 82)$}{&$ 987( 64)$}{&$1028( 47)$}{&$1030( 47)$}}%
{\selectcols{&$ 729(119)$}{&$ 814( 92)$}{&$ 842( 81)$}{&$ 843( 81)$}{&$ 914( 63)$}{&$ 956( 46)$}{&$ 958( 46)$}}\\
 \putrow{7.0                                               }{&$0.96E-02$}%
{\selectcols{&$ 817(109)$}{&$ 888( 82)$}{&$ 912( 73)$}{&$ 913( 73)$}{&$ 972( 66)$}{&$1007( 61)$}{&$1008( 62)$}}%
{\selectcols{&$ 713(107)$}{&$ 784( 81)$}{&$ 807( 71)$}{&$ 808( 71)$}{&$ 868( 64)$}{&$ 902( 60)$}{&$ 904( 60)$}}\\
 \putrow{10.0                                              }{&$0.67E-01$}%
{\selectcols{&$ 859(149)$}{&$ 908(109)$}{&$ 925( 97)$}{&$ 926( 96)$}{&$ 967( 85)$}{&$ 991( 84)$}{&$ 992( 84)$}}%
{\selectcols{&$ 746(147)$}{&$ 795(107)$}{&$ 812( 95)$}{&$ 812( 95)$}{&$ 854( 83)$}{&$ 878( 82)$}{&$ 879( 82)$}}\\
\noalign{\hrule}
 \putrow{}{&}{\selectcols{&}{&}{&}{&}{&}{&}{&}}{\selectcols{&}{&}{&}{&}{&}{&}{&}}\\[-10pt]
 \putrow{\ensuremath{\overline\Lambda}                     }{&          }%
{\selectcols{&$ 895(218)$}{&$ 926(148)$}{&$ 937(130)$}{&$ 937(130)$}{&$ 963(107)$}{&$ 978(112)$}{&$ 978(113)$}}%
{\selectcols{&$ 714(179)$}{&$ 751(125)$}{&$ 763(111)$}{&$ 764(110)$}{&$ 796( 94)$}{&$ 815( 95)$}{&$ 816( 95)$}}\\
 \putrow{\ensuremath{-\lambda_1}                      }{&          }%
{\selectcols{&$-1.7(3.4)$}{&$-0.8(2.2)$}{&$-0.5(1.8)$}{&$-0.5(1.8)$}{&$ 0.2(1.2)$}{&$ 0.7(1.4)$}{&$ 0.8(1.4)$}}%
{\selectcols{&0.5(2.0)}{&1.2(1.3)}{&1.4(1.1)}{&1.4(1.1)}{&2.0(0.7)}{&2.3(0.9)}{&2.4(0.9)}}\\
 \noalign{\hrule}
 \end{tabular}
 \end{center}
 \endgroup
 \par\vfil\penalty-5000\vfilneg

\caption{Binding energies in MeV for the $\Lambda_h$ baryon.}
\label{tab:LambdabarL}
\end{table}

To remove the uncertainty in $\overline{\Lambda}$ and $\lambda_1$ 
due to the perturbative estimate of $E_0$ we construct differences
of binding energies in which $E_0$ drops out. The intercept of
a linear fit to the spin-averaged $\Lambda_h-H$ and $\Sigma_h-\Lambda_h$ 
splittings versus $1/\overline{M}$ gives 
%
\ba \overline{\Lambda}({\Lambda_b}) - \Lambdabar(B) &=& 
        415(156)({}^{+17}_{-25})\MeV, \nonumber \\
     \Lambdabar(\Sigma_b) - \Lambdabar(\Lambda_b) &=&
        176(152)({}^{+9}_{-10}) \MeV. 
\ea 
In both cases we find no significant dependence on $1/\overline{M}$.
This suggests that the corresponding $\lambda_1$ are roughly the same.
A similar construction for states with different light quarks gives:
\ba
\Lambdabar({B_s}) - \Lambdabar({B_d}) &=&
    81(31)({}^{-3}_{+5})(^{+18}_{-0}) \MeV, \nonumber \\
\lambda_1({B_s}) - \lambda_1({B_d}) &=& -0.10(28)(^{+2}_{-0}) {\rm GeV}^2\,.
\ea

Lastly, we estimate $\lambda_2$ from the slope of the hyperfine
splitting calculated in Sec.~\ref{ss:BstarB}. We
find 
\ba
\lambda_2(B_d) &=& 0.069(19)({}^{+6}_{-8}){\rm GeV}^2\,, \nonumber \\
\lambda_2(B_s) &=& 0.078(12)({}^{+6}_{-8})({}^{+2}_{-0}){\rm GeV}^2\,.
\ea

These parameters have previously been calculated by the Rome collaboration 
using HQET \cite{Mb97HQET,HQET96Rome}.  They find 
\begin{eqnarray}
  \overline{\Lambda}(B) = 180({}^{+30}_{-20})  {\rm MeV};&&\qquad
  -\lambda_1({B_d})     = -0.09(14) {\rm GeV}^2;           \nonumber \\
  \lambda_1({B_s})-  \lambda_1({B_d}) = -0.09(4) {\rm GeV}^2;&&\qquad
  \hphantom{-} \lambda_2(B_d)        = 0.070(15) {\rm GeV}^2 \,.
\end{eqnarray}
It is important to note that their definition of $ \overline{\Lambda}$
and $\lambda_1$ includes a non-perturbative subtraction of the
ultra-violet divergence. Thus, the only results that can be compared
directly are those for $\lambda_1({B_s})- \lambda_1({B_d})$ and $
\lambda_2(B_d)$. The experimental values for these two quantities are
\begin{eqnarray}
\lambda_1(B_s) - \lambda_1(B_d) = {} &\displaystyle 2
        \frac{({\overline M}_{B_s} - {\overline M}_{B}) -
              ({\overline M}_{D_s} - {\overline M}_{D})}
             {1/{\overline M}_{D} - 1/{\overline M}_B} &{} = -0.06(2) {\rm GeV}^2, \nonumber\\
\lambda_2(B_d) ={} & \displaystyle \frac{M_{B_d^*}^2 - M_{B_d}^2} 4 &{} = \hphantom{-} 0.12(1) {\rm GeV}^2.
\end{eqnarray}

\section{Conclusions}
\label{sec:conclusions}

We have presented an analysis of heavy-light mesons and baryons using
a non-relativistic formulation (NRQCD) for the bottom quark. Estimates
of meson masses with one $b$ quark and baryons with one or two $b$
quarks are given in Tables~\ref{tab:meson_summary} and
\ref{tab:baryexp}.  Using the $B$ meson to fix the $b$ quark mass, we
estimate $m_{\overline{MS}} (m_{\overline{MS}}) = 4.35(10)({}^{-3}_{+2})(20)$ GeV.
This is consistent with previous lattice determinations of $m_b$
using the $\Upsilon$ binding energy \cite{Mb94NRQCD,kent,PDB98}, or
HQET \cite{PDB98,Mb97HQET,APE98}.  A more direct comparison will be possible
after we extract, using the same set of lattices and propagators,
$m_b$ from the $\Upsilon$ binding energy. 

A significant feature of our calculation is that we can resolve the
$P$ states.  We find that $M_{B^\ast_0} < M_{B^\ast_2}$.
Using the interpolating operators based on the $LS$
coupling scheme, we could not distinguish between the $1^+$ and
$1^{+\prime}$ states, as these mix.  Also, we resolve $b$ baryon
hyperfine splittings for the first time on the lattice.

The mass splittings are analyzed in terms of a qualitative picture
based on a non-relativistic quark model that is described in
Sec.~\ref{sec:mesons}. We find that the dependence of the
splittings on the light and heavy quark masses are in agreement with
this picture.  Quantitatively, the radial ($2S-1S$), orbital ($P-S$),  
$\overline{\Sigma} - \Lambda$, and $\Lambda - \overline{B}$ splittings are 
found to be within $1 \sigma$ ($\sim 20\%$) of the experimental values. 

We are able to resolve hyperfine splittings in both mesons and
baryons.  The most significant difference from experimental numbers is
in the $B^\ast - B$ hyperfine splitting.  Such an underestimate of
hyperfine splittings is a general feature of quenched calculations
(light-light, heavy-light and heavy-heavy). Another uncertainty
associated with the quenched approximation is in fixing the strange
quark mass.  As a result, splittings which are sensitive to the light
quark mass have an uncertainty of up to roughly $20\%$ when extrapolated to
the strange quark mass.

We have calculated the HQET parameters $\overline\Lambda$,
$\lambda_1$, and $\lambda_2$ for both the $B$ and $\Lambda_b$.
$\overline\Lambda$ and $\lambda_1$ have large uncertainties due to the
perturbative determination of the shift in the energy of the heavy
quark, $E_0$.  The differences in these quantities between different
hadrons do not have this ambiguity and are, therefore, much better
determined.

\vspace{.2in}
\noindent
\underline{Acknowledgements}

\vspace{.1in}
\noindent
This work has been supported by grants from the US Department of
Energy, (DE-FG02-91ER40690, DE-FG03-90ER40546, DE-FG05-84ER40154,
DE-FC02-91ER75661, DE-LANL-ERWE161), NATO (CRG 941259), and PPARC.
A.~Ali~Khan thanks the Graduate
School of the Ohio State University for a University Postdoctoral
Fellowship, and the Research for the Future Program of 
the Japanese Society for the Promotion of Science.  
S.~Collins is grateful for funding from the Alexander von
Humboldt foundation and the Royal society of Edinburgh.  C. Davies
thanks the Institute for Theoretical Physics, Santa Barbara, for
hospitality and the Leverhulme Trust and Fulbright Commission for
funding while this work was being completed.  J. Sloan would like to
thank the Center for Computational Sciences, University of Kentucky,
for support.  We thank J. Hein for discussions. 
The simulations reported here were done on CM5's at the
Advanced Computing Laboratory at Los Alamos under a DOE Grand
Challenges award, and at NCSA under a Metacenter allocation.


\ifx\href\undefined\def\href#1#2{{#2}}\fi
\def\spireshome{http://www.slac.stanford.edu/cgi-bin/spiface/find/hep/www?FORMAT=WWW&}
\def\xxxhome{http://xxx.lanl.gov/abs/}
{\catcode`\%=12
\xdef\spiresjournal#1#2#3{\noexpand\protect\noexpand\href{\spireshome
                          rawcmd=find+journal+#1%2C+#2%2C+#3}}
\xdef\spireseprint#1#2{\noexpand\protect\noexpand\href{\spireshome rawcmd=find+eprint+#1%2F#2}}
\xdef\spiresreport#1{\noexpand\protect\noexpand\href{\spireshome rawcmd=find+rept+#1}}
\xdef\spireskey#1{\noexpand\protect\noexpand\href{\spireshome key=#1}}
\xdef\xxxeprint#1{\noexpand\protect\noexpand\href{\xxxhome #1}}
}
\def\eprint#1#2{\xxxeprint{#1/#2}{#1/#2}}
\def\report#1{\spiresreport{#1}{#1}}
\def\nohref{}

\makeatletter
\def\putpaper{\@ifnextchar [{\@putpaper}{\@putpaper[]}}
\def\@putpaper[#1]{\edef\refpage{\zeros\the\count0}%
              \def\nohref{}%
              {\def\ {+}\def\nohref##1{}\edef\temp{\ifx\relax#1\relax
               \noexpand\spiresjournal{\journalname}{\volume}{\refpage}%
               \else\noexpand\xxxeprint{#1}\fi}\expandafter}\temp
               {\sfcode`\.=1000{\journalname} \journalformat}\egroup}
\def\putpage{\@ifnextchar [{\@putpage}{\@putpage[]}}
\def\@putpage[#1]{\edef\refpage{\the\count0}%
              \def\nohref{}%
              {\def\ {+}\def\nohref##1{}\edef\temp{\ifx\relax#1\relax
               \noexpand\spiresjournal{\journalname}{\volume}{\refpage}%
               \else\noexpand\xxxeprint{#1}\fi}\expandafter}\temp
              {\refpage}\egroup}
\def\dojournal#1#2 (#3) {\def\zerocommand{\@dojournal{#1}{#2} ({#3})}
                        \def\zeros{}\storezeros}
\def\@dojournal#1#2 (#3){\def\journalname{#1}\def\volume{#2}%
                          \def\refyear{#3}\afterassignment\putpaper\bgroup
                          \count0=}
\def\storezeros{\futurelet\zerocheck\@storezeros}
\def\@storezeros#1{\ifx0\zerocheck\edef\zeros{\zeros0}\expandafter\storezeros
                   \else\zerocommand#1\fi}
\def\morepage{\let\zerocommand=\@morepage\def\zeros{}\storezeros}
\def\@morepage{\afterassignment\putpage\bgroup\count0=}
\def\supresslink{\def\spiresjournal##1##2##3{}}

\def\APNY#1{\dojournal{Ann.\ Phys.\ \nohref{(N.\ Y.)}}{#1}}
\def\CMP#1{\dojournal{Comm.\ Math.\ Phys.}{#1}}
\def\IJMPC#1{\dojournal{Int.\ J.\ Mod.\ Phys.}{C#1}}
\def\IJMPE#1{\dojournal{Int.\ J.\ Mod.\ Phys.}{E#1}}
\def\JAP#1{\dojournal{J.\ App.\ Phys.}{#1}}

\def\MPA#1{\dojournal{Mod.\ Phys.\ Lett.}{A#1}}
\def\MPLA#1{\dojournal{Mod.\ Phys.\ Lett.}{A#1}}
\def\NP#1{\dojournal{Nucl.\ Phys.}{B#1}}
\def\NPA#1{\dojournal{Nucl.\ Phys.}{A#1}}
\def\NPB#1{\dojournal{Nucl.\ Phys.}{B#1}}
\def\NPBPS#1{\dojournal{Nucl.\ Phys.\ B \nohref(Proc.\ Suppl.\nohref)}{\nohref #1}}
\def\NPAPS#1{\dojournal{Nucl.\ Phys.\ \nohref(Proc.\ Suppl.\nohref)}{\nohref A#1}}
\def\NC#1{\dojournal{Nuovo Cimento }{#1}}
\def\PRL#1{\dojournal{Phys.\ Rev.\ Lett.}{#1}}
\def\PR#1{\dojournal{Phys.\ Rev.}{#1}}
\def\PRep#1{\dojournal{Phys.\ Rep.}{ #1}}
\def\PRB#1{\dojournal{Phys.\ Rev.}{B #1}}
\def\PRC#1{\dojournal{Phys.\ Rev.}{C #1}}
\def\PRD#1{\dojournal{Phys.\ Rev. \ D}{ #1}}
\def\PRE#1{\dojournal{Phys.\ Rev.}{E #1}}
\def\PL#1{\dojournal{Phys.\ Lett.}{#1B}}
\def\PLA#1{\dojournal{Phys.\ Lett. \ A}{#1}}
\def\PLB#1{\dojournal{Phys.\ Lett. \ B}{#1}}
\def\RMP#1{\dojournal{Rev.\ Mod.\ Phys.}{#1}}
\def\PREP#1{\dojournal{Phys.\ Rep.}{#1}}
\def\ZEITC#1{\dojournal{Z.\ Phys.}{C#1}}
\def\ZPC#1{\dojournal{Z.\ Phys.}{C#1}}

\def\ie{{\sl i.e.}}
\def\etal{{\it et al.}}
\def\etc{{\it etc.}}
\def\ibid{{\it ibid}}


\let\super=^
\catcode`\^=13
\def^{\ifmmode\super\else\initialsep\fi}

\def\journalformat{{\bf \volume}, \refpage\ (\refyear)}
\def\initialsep{}

\end{document}